\documentclass[twocolumn,aps,pre,preprintnumbers,amsmath,amssymb,nofootinbib]{revtex4-1}

\usepackage[dvipdfmx]{graphicx}
\usepackage{dcolumn}
\usepackage{bm}
\usepackage{textcomp}
\usepackage{booktabs}
\usepackage{color}

\begin{document}

\title{Coarse-grained molecular dynamics simulation \\ for uptake of nanoparticles\\ 
into a charged lipid vesicle dominated by electrostatic interactions} 

\author{Naofumi Shimokawa}
\email{nshimo@jaist.ac.jp}
\affiliation{%
School of Materials Science, Japan Advanced Institute of Science and Technology, Ishikawa 923-1292, Japan
}%
\author{Hiroaki Ito}
\altaffiliation[Present address: ]{Department of Physics, Chiba University, Chiba 263-8522, Japan
}%
\affiliation{%
Department of Mechanical Engineering, Graduate School of Engineering, Osaka University, Osaka 565-0871, Japan
}%

\author{Yuji Higuchi}
\affiliation{%
Institute for Solid State Physics, University of Tokyo, Chiba, 227-8581, Japan
}%

\date{\today}

\begin{abstract}
We use a coarse-grained molecular dynamics simulation to investigate the interaction between neutral or charged nanoparticles (NPs) and a vesicle consisting of neutral and negatively charged lipids.
We focus on the interaction strengths of hydrophilic and hydrophobic attraction and electrostatic interactions between a lipid molecule and an NP.
A neutral NP passes through the lipid membrane when the hydrophobic interaction is sufficiently strong.
As the valence of the positively charged NP increases, the membrane permeation speed of the NP is increased compared with the neutral NP and charged lipids are accumulated around the charged NP.
A charged NP with a high valence passes through the lipid membrane via a transient channel formed by charged lipids or transportlike endocytosis.
These permeation processes can be classified based on analyses of the density correlation function.
When the non-electrostatic interaction parameters are large enough, a negatively charged NP can be adsorbed on the membrane and a neutral lipid-rich region is formed directly below the NP.
The NP is spontaneously incorporated into the vesicle under various conditions and the incorporation is mediated by the membrane curvature.
We reveal how the NP's behavior depends on the NP valence, size, and the non-electrostatic interaction parameters.
\end{abstract}

\maketitle

\section{INTRODUCTION}
\label{intro}
Phospholipids, which are typical amphiphilic molecules, self-assemble into a bilayer structure in the aqueous phase so as to protect their hydrophobic parts from the aqueous phase.
The bilayer structure is essentially the same as the structure of the plasma membrane, and thus phospholipid vesicles, which are spherical closed bilayers, are widely used as a simple model for cells.
Various physicochemical phenomena in lipid vesicles are deeply related to biocellular functions~\cite{KA}.
In multicomponent lipid membranes, a compositional heterogeneous structure emerges spontaneously at lower temperatures~\cite{Keller1,Keller2}.
This phase separation has attracted great attention in relation to raft domains in biomembranes~\cite{raft,Simons}.
Moreover, dynamic membrane deformation, such as endocytosis and exocytosis, can be reproduced in some vesicular systems~\cite{Hamada}.

Biomembranes also contain negatively charged lipids in addition to zwitterionic lipids.
Negatively charged lipids, such as phosphatidylserine, phosphatidylglycerol, phosphatidylinositol, phosphatidic acid, and cardiolipin, play important roles in protein adsorption on membranes~\cite{Ben-Tal} and channel activity~\cite{Hille}.
In addition, several studies reported that negatively charged lipids greatly affect phase separation and membrane deformation in artificial lipid membranes.
For example, the suppression of the phase separation in unsaturated charged lipid-containing membranes was examined, and the addition of salt made the phase behavior similar to that of neutral membranes~\cite{Shimokawa1,Dimova1,Himeno1,Dimova2}.
This behavior was explained by theoretical and numerical studies~\cite{May,Shimokawa2,Shimokawa3,Ito}.
In addition, decreasing the temperature spontaneously changed the vesicle morphology from spherical to disk shape~\cite{Himeno2}.
Thus, electric charges on lipids play important roles in membrane stability, which is a crucial factor in the uptake of proteins and drugs into cells.

The interaction between nanoparticles (NPs) and the plasma membrane is important in NP cytotoxicity in the development of nanotechnology and drug delivery~\cite{Albanese,Jaggers}.
Moreover, this type of interaction is also relevant to the uptake of viruses into cells~\cite{virus}.
Many studies using living cells have been performed to reveal how NPs and nanomaterials are incorporated into cells and the cytotoxicity caused by NP incorporation.
Various materials, such as fullerene~\cite{Rouse}, carbon nanotubes~\cite{Shvedova}, dendrimers~\cite{Seib,Albertazzi}, gold NPs~\cite{Chithrani}, cobalt oxide particles~\cite{Bossi}, and quantum dots~\cite{Ruan}, have been investigated.
Studies using vesicles focus on the interaction between NPs and lipid membranes and the uptake behavior of NPs into vesicles~\cite{Chen}.
Changes in membrane surface properties~\cite{Granick,Wei}, membrane protrusion~\cite{Li}, and emergence of lipid accumulation~\cite{Hinderliter,McLaughlin,Scheve} have been observed by NP and protein adsorption on lipid membranes.
Furthermore, the uptake behavior of NPs into vesicles has also been studied; the membrane permeation and endocytic behavior of NPs or nanoemulsions~\cite{Banerji,Bihan,Miyake} and the effects of phase-separated domains on NP permeation~\cite{Takeuchi} have been examined experimentally.

One of the NP uptake behaviors is a passive process.
Passive processes are caused by the interaction between the NP and membrane without an external energy supply.
To understand NP uptake better, it is important to understand the passive process mechanism, which is the basic uptake mechanism ~\cite{Vacha}.
Because it is difficult to observe the dynamic behavior of NP uptake on the nanometer scale experimentally, numerical simulation studies have been performed.
The hydrophobic interaction between the lipid membrane and the hydrophobic part of NPs, the specific interaction between the lipid membrane and ligands grafted onto the surface of NPs, and the electrostatic interactions between the polar groups on the lipid headgroups and charged NPs all play important roles in NP uptake.
In the studies of the hydrophobic interactions and ligand interactions, the size, shape, and surface properties of the NPs~\cite{Vacha,Gompper,Lehn1,Lehn2,Ye_Li} and the cooperative behavior of multiple NPs~\cite{Yue} have been discussed.
In particular, the receptor-mediated endocytosis of NPs can be described by the competition between the curvature elastic energy of lipid membrane and the attractive energy between lipid membrane and NPs via ligand-receptor binding~\cite{Vacha}.
The endocytic motion of larger NPs occurs easily, since the energy loss by curvature change becomes smaller and the energy gain by the attractive energy becomes larger.
However, the model is not sufficient because the lipid membrane is consisting of single component and the long-range interaction such as electrostatic interaction is not taken into account.

Although calculations that include the electrostatic interactions usually become complicated due to the long-range nature of electrostatic interactions, several studies have simulated NP uptake mediated by electrostatic interactions~\cite{Yang_Li,Liu}.
In particular, the interaction between charged NPs and a mixed membrane consisting of neutral and charged lipids was investigated to mimic the cellular environment~\cite{Yang_Li,Li_Gu,Nangia}.
Lipids with charge opposite to NPs are accumulated in the NP adsorption part~\cite{Yang_Li,Li_Gu}.
As a result, the large adsorption energy derived from electrostatic interaction is obtained, which overcomes the curvature energy loss, and endocytic movement occurs.
This picture is essentially the same as the receptor-mediated endocytosis. 
However, there are still few of these studies and most studies use a box-spanning flat membrane, where the asymmetry between the inner and outer space is not incorporated, and the vesicular environment with electrostatic interactions is usually not considered.
Furthermore, it is necessary to investigate how NP incorporation into a vesicle depends on interaction and NP size because the types of NPs and drugs are diverse.

We have previously conducted a coarse-grained molecular dynamics (CGMD) simulation to examine the dynamic behavior of binary lipid membranes consisting of neutral and charged lipids~\cite{Ito}.
In the present study, we investigate the dynamic behavior of an NP placed near the outer leaflet of a charged lipid membrane based on the previous CGMD simulation.
CGMD is suitable for describing the uptake of NPs into a vesicle while changing the interaction strength.
We systematically change the strengths of hydrophilic, hydrophobic, and electrostatic interactions between a lipid molecule and an NP.
The details of the CGMD simulation are introduced in Sec.~\ref{methods}. 
The interaction between a neutral NP and a charged membrane and the dynamic behavior of a neutral NP are presented in Sec.~\ref{neutral}.
The interaction and the dynamic behavior of a charged NP are discussed in Sec.~\ref{charged}.
Finally, we discuss the implications of our work and compare it with other studies in Sec.~\ref{discussion}.

\section{METHODS}
\label{methods}

Based on our previous CGMD simulation~\cite{Ito}, which was based on the CGMD simulation by Cooke {\it et al.}~\cite{Cooke}, we calculate the behavior of charged lipid membranes.
This model describes the phase separation and spontaneous membrane morphological change of charged lipid membranes well~\cite{Ito, Himeno2}.
In the present study, we calculate the interaction between a charged lipid vesicle modeled by our previous study and an NP.

A lipid molecule consists of one hydrophilic head bead and two hydrophobic tail beads. 
For convenience, we refer to the hydrophilic bead, the hydrophobic bead in the center of lipid molecule, and the hydrophobic bead at the end of lipid molecule as a-bead, b-bead, and c-bead, respectively.
These beads are connected linearly through springs.
The lipid-lipid and NP-lipid interactions are indicated by superscript LL and NL, respectively.
The excluded-volume interaction between two beads separated by distance $r$ is
\begin{equation}
\label{rep}
V^{\rm (LL)}_{\rm re}(r;b)=\begin{cases}
                  4v\left[ \left( \frac{b}{r} \right)^{12} - \left( \frac{b}{r} \right)^{6} + \frac{1}{4} \right], & r \leq r_{\rm c}, \\
                  0, & r>r_{\rm c},
                 \end{cases}
\end{equation}
where $r_{\rm c}=2^{1/6}b$. 
$v=k_{\rm B}T$ is the unit of energy, where $k_{\rm B}$ is the Boltzmann constant and $T$ is absolute temperature.
To form a stable bilayer, $b_{\rm head,head}=b_{\rm head,tail}=0.95 \sigma$ and $b_{\rm tail,tail}=\sigma$ are chosen~\cite{Ito,Cooke}. 
Here, $\sigma$ is the unit of length corresponding to the cross-sectional diameter of a single lipid molecule.
The stretching and bending potentials of bonds between connected beads are expressed as
\begin{equation}
\label{bond}
V_{\rm bond}(r)=\frac{1}{2}k_{\rm bond}(r-\sigma)^{2}
\end{equation}
and
\begin{equation}
\label{bend}
V_{\rm bend}(\theta)=\frac{1}{2}k_{\rm bend}(1-\cos \theta)^{2},
\end{equation}
where $k_{\rm bond}=500v$ is the bonding strength of connected beads, $k_{\rm bend}=60v$ is the bending stiffness of a lipid molecule, and $\theta$ is the angle between adjacent bond vectors.
The hydrophobic attractive interaction among hydrophobic beads is expressed as
\begin{equation}
\label{attr}
V^{\rm (LL)}_{\rm at}(r)=\begin{cases}
                -v, & r<r_{\rm c}, \\
                -v \cos^{2} \left[\frac{\pi(r-r_{\rm c})}{2 w_{\rm c}} \right], & r_{\rm c} \leq r \leq r_{\rm c}+w_{\rm c}, \\
                0, & r>r_{\rm c}+w_{\rm c},
                 \end{cases}
\end{equation}
where $w_{\rm c}$ is the cut-off length for the attractive potential. 
The electrostatic repulsion between charged hydrophilic beads is described as the Debye-H\"{u}ckel potential
\begin{equation}
\label{elec}
V^{\rm (LL)}_{\rm el}(r)=v \ell_{\rm B} z_{1}z_{2}\frac{\exp(-r/\ell_{\rm D})}{r},
\end{equation}
where $\ell_{\rm B}=\sigma$ is the Bjerrum length, $z_{1}$ and $z_{2}$ are the valencies of the interacting charged headgroups,
and $\ell_{\rm D}=\sigma \sqrt{\epsilon k_{\rm B}T/n_{0}e^{2}}$ is the Debye--H\"{u}ckel screening length, which is related to the bulk salt concentration, $n_{0}$, where $\epsilon$ and $e$ are the dielectric constant of the solution and elementary charge, respectively. 
As a salt, we consider a symmetric monovalent salt, such as NaCl.
Because typical charged headgroups, such as phosphatidylglycerol, phosphatidylserine, phosphatidylinositol, and phosphatidic acid, have a monovalent charge, we set $z_{1}=z_{2}=-1$. 
We do not set a cutoff for this screened electrostatic interaction.
The modeling so far is the same as our previous study~\cite{Ito}.

Next, the interaction between an NP with diameter $d$ and lipids is introduced.
The potential between an NP and lipid beads is expressed based on the Lennard--Jones potential,
\begin{equation}
\label{cllj}
V^{\rm (NL)}_{\rm LJ}(r)=4v_{\rm b}\left[ \left( \frac{b}{r} \right)^{12} - \left( \frac{b}{r} \right)^{6} \right]. 
\end{equation}
We set $v_{\rm b}=v_{\rm nh}$ for the hydrophilic beads and  $v_{\rm b}=v_{\rm nt}$ for the hydrophobic beads.
The electrostatic interaction between a charged NP and charged beads is written as
\begin{multline}
\label{clel}
V^{\rm (NL)}_{\rm el}(r)=v\frac{\ell_{\rm B} z_{1} z_{\rm n}}{[1+(\sigma/2\ell_{\rm D})][1+(d/2\ell_{\rm D})]} \\ 
\frac{\exp\{-[r-(\sigma+d)/2]/\ell_{\rm D}\}}{r},
\end{multline}
where $z_{\rm n}$ is the NP valence.
We consider that the electric charges are distributed on the surface of NP in this potential~\cite{Sakaue}.
Our simulation implicitly considers the same salt solution for exterior and interior of the vesicle and the same salt solution is also present inside the membrane in this modeling of electrostatic interaction.
It should be noted that we do not consider the presence of water molecules explicitly.
Instead, it is included as electrostatic interaction implicitly.
Therefore, in this calculation, volume change of water phase inside the vesicle and osmotic pressure are not taken into account.

Each bead and an NP obey the stochastic dynamics described by the Langevin equation
\begin{equation}
\label{langevin}
m\frac{d^{2}\bm{r}_{i}}{dt^{2}}=-\eta\frac{d\bm{r}_{i}}{dt}+\bm{f}^{V}_{i}+\bm{\xi}_{i},
\end{equation}
where $m=1$ and $\eta=1$ are the mass and drag coefficients, respectively.
Forces $\bm{f}^{V}$ are obtained from the derivatives of the interaction potentials Eqs. (\ref{rep})--(\ref{clel}) for the beads and Eqs. (\ref{cllj}) and (\ref{clel}) for the NPs. 
Constant $\tau=\eta\sigma^{2}/v$ is chosen as the unit for the time scale, and the time increment is set at $dt=7.5 \times 10^{-3}\tau$ which is sufficiently small to keep the stability of the calculation.
The Brownian force, $\bm{\xi}_{i}$, satisfies the fluctuation-dissipation theorem
\begin{equation}
\label{flu-dis}
\left< \bm{\xi}_{i}(t)\bm{\xi}_{j}(t')\right>=6v\eta\delta_{ij}\delta(t-t').
\end{equation}

The calculated lipid membrane is a binary mixture of A-lipids with a monovalent electric charge at the head beads, corresponding to charged lipids, and neutral lipid (B-lipid)s. 
Charged lipid (A-lipid) is represented by red hydrophilic bead and green hydrophobic beads, while neutral lipid (B-lipid) is represented by yellow hydrophilic bead and cyan hydrophobic beads.
A lipid vesicle is composed of 2500 charged lipids (A-lipids) and 2500 neutral lipids (B-lipids), and the salt concentration is 100 mM.
The fraction of charged lipids in plasma membranes is roughly 20--30\%.
On the other hand, various compositions have been used in the experiments with artificial lipid membranes to investigate the effect of electrostatic interaction~\cite{Shimokawa1,Dimova1,Himeno1,Dimova2,Himeno2}.
Hence, we calculate 50\% charged lipid-containing membranes to see the effect of the electrostatic interaction clearly.
We constructed the vesicles with setting lipids randomly by hand as spheres as shown in Fig. S1(a)~\cite{SI} and then started the simulations~\cite{Himeno2}.
Before we calculate the interaction between an NP and the lipid membrane, we perform the calculation until $t=10 \times 7500\tau$ to relax the lipid membrane structure sufficiently~\cite{Ito}.
The cutoff lengths for the attractive interaction among hydrophobic beads are set to $w_{\rm c}^{\rm AA}/\sigma=1.8$, $w_{\rm c}^{\rm BB}/\sigma=1.7$, and $w_{\rm c}^{\rm AB}/\sigma=1.575$.
When the phase separation takes place at $w_{\rm c}^{\rm AB}/\sigma=1.55$, the attractive energy among hydrophobic beads $V^{\rm (LL)}_{\rm at}$ gradually decreases and reaches the plateau after $t=2 \times 7500 \tau$ as shown in Fig. S1(b)~\cite{SI}.
Therefore, the membrane structure at $t=10 \times 7500 \tau$ can be regarded as a sufficiently relaxed state, even if the phase separation does not occur at $w_{\rm c}^{\rm AB}/\sigma=1.575$.
More details of the relaxation process are shown in Fig. S1~\cite{SI}.
As the initial state of the membrane structure, we choose a sufficiently relaxed membrane at $t=10 \times 7500\tau$.
After an NP is put on the outer leaflet of a vesicle with zero velocity (we reset this time as $t=0 \times 7500\tau$), we start the calculation for $t=1 \times 7500\tau$.
As for NP size, it has been reported that NPs less than 3.1 nm can easily pass through a lipid membrane~\cite{Takeuchi}.
Therefore, we use two sizes of NPs, $d/\sigma=2$ and $5$ which roughly correspond to 1.4 nm and 3.5 nm, respectively.
We use $v_{\rm nh}/k_{\rm B}T$=1, 2, and 3, $v_{\rm nt}/k_{\rm B}T$=1, 2, and 3, and $z_{\rm n}=0, +10, +50, +150, +200,$ and $-50$.
$v_{\rm nh}$ and $v_{\rm nt}$ are related to the phase equilibrium between adsorbed and dispersed states, where the solubility is roughly estimated as $\exp(-v^{*}/k_{\rm B}T)$ for $v^{*} = v_{\rm nh}$ or $v_{\rm nt}$.
The all parameters are summarized in Table S1~\cite{SI}.
Calculations are performed at least three times for each condition to ensure data reproducibility.
We show a representative result in the calculations we performed.
The results shown are almost identical to the other calculation results which are shown in Figs. S6, S7, S8, S9, and S10~\cite{SI}.

\section{RESULTS}

\subsection{Neutral NP}
\label{neutral}

In this section, we investigate the interaction between a neutral NP ($z_{\rm n}=0$) and a charged vesicle.
The non-electrostatic interaction parameters of the Lennard--Jones potential $V^{\rm (NL)}_{\rm LJ}$, $v_{\rm nh}/k_{\rm B}T$ and $v_{\rm nt}/k_{\rm B}T$, are changed from
$1$ to $3$, and the dynamic behavior of an NP with the diameter $d/\sigma=5$ or $2$ is examined.
To ensure data reproducibility, the results obtained by the other two calculations are shown in Fig. S6~\cite{SI}.

\begin{figure*}[t!]
\begin{center}
\includegraphics{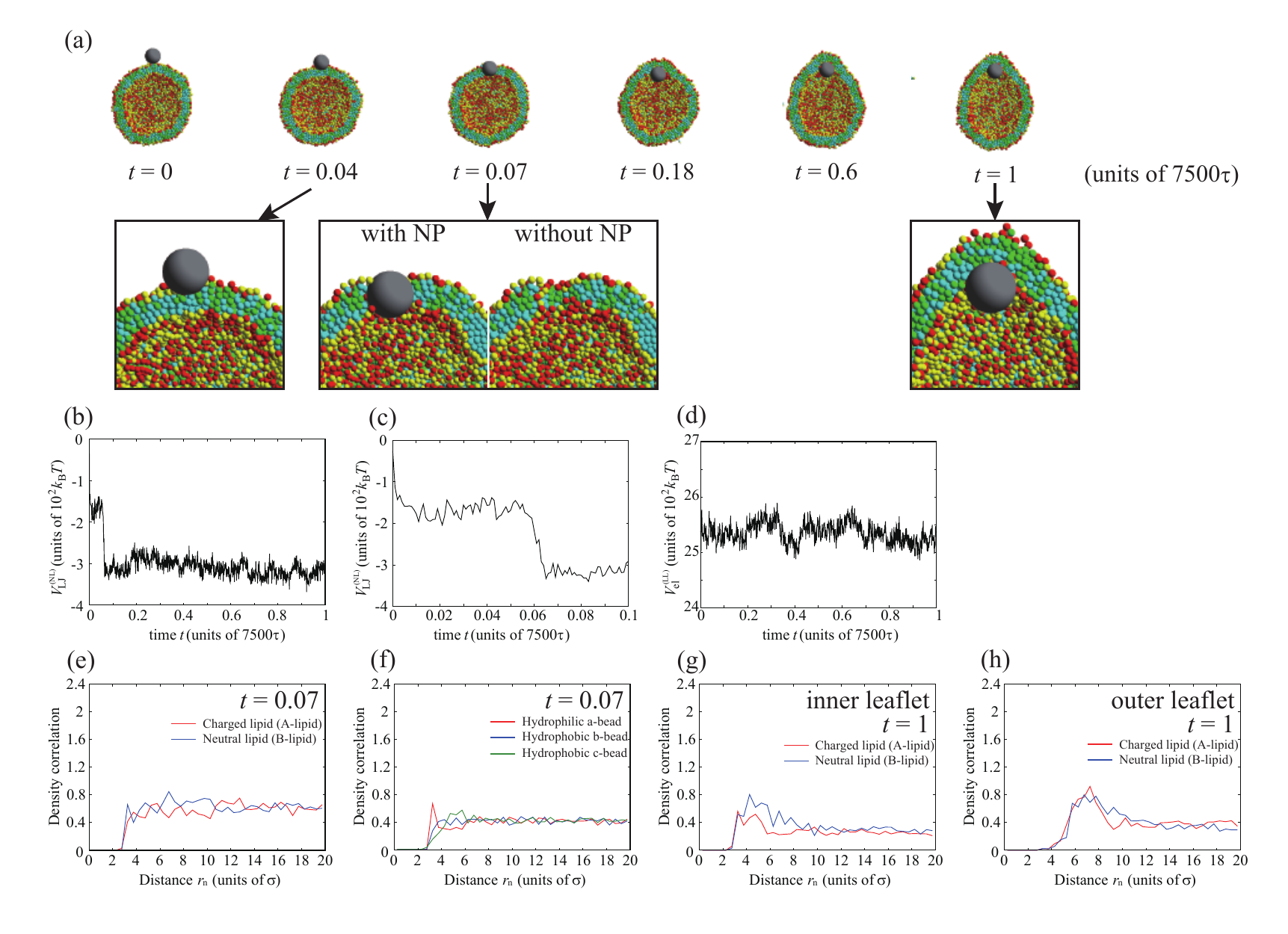}
\end{center}
\caption{
(a) Typical snapshots of the permeation sequence of a neutral NP ($z_{\rm n}=0$) for $d/\sigma=5$, $v_{\rm nh}/k_{\rm B}T=2$, and $v_{\rm nt}/k_{\rm B}T=3$.
Charged lipid (A-lipid) is represented by red hydrophilic beads and green hydrophobic beads, while neutral lipid (B-lipid) is represented by yellow hydrophilic beads and cyan hydrophobic beads.
(b) Time evolution of the Lennard--Jones potential between the NP and lipids, $V^{\rm (NL)}_{\rm LJ}$, for $t=1 \times 7500\tau$.
(c) Enlarged graph for time evolution of the Lennard--Jones potential between the NP and lipids, $V^{\rm (NL)}_{\rm LJ}$ until $t=0.1 \times 7500\tau$.
(d) Time evolution of the electrostatic interaction between lipids, $V^{\rm (LL)}_{\rm el}$, for $t=1 \times 7500\tau$.
(e,f) Density correlation functions of lipids and beads around the NP at $t=0.07 \times 7500\tau$, respectively.
Red and blue lines in (e) indicate the density correlation functions of A- and B-lipids, respectively.
Red, blue, and green lines in (f) indicate the density correlation functions of a-, b-, and c-beads, respectively.
(g,h) Density correlation functions of A- and B-lipids in the inner leaflet in (g) and outer leaflet in (h) around the NP at $t=1 \times 7500\tau$.
Red and blue lines indicate the density correlation functions of A- and B-lipids, respectively.
}
\label{fig1}
\end{figure*}

Figure~\ref{fig1} shows typical dynamic behavior of an NP for $d/\sigma=5$, $v_{\rm nh}/k_{\rm B}T=2$, and $v_{\rm nt}/k_{\rm B}T=3$, and the NP passes through the lipid membrane (Fig.~\ref{fig1}(a)).
First, the NP is strongly adsorbed on the outer leaflet (snapshot at $t=0.04 \times 7500 \tau$ in Fig.~\ref{fig1}(a)), passes through the hydrophobic region of the lipid membrane  (snapshot at $t=0.07 \times 7500 \tau$ in Fig.~\ref{fig1}(a)), and is finally adsorbed on the inner leaflet (snapshots after $t=0.18 \times 7500 \tau$ in Fig.~\ref{fig1}(a)).
The Lennard--Jones potential between the NP and lipids, $V^{\rm (NL)}_{\rm LJ}$, is shown in Fig.~\ref{fig1}(b).
Figure~\ref{fig1}(c) shows the enlarged view from $t=0$ to $0.1\times 7500\tau$ in Fig.~\ref{fig1}(b).
The potential $V^{\rm (NL)}_{\rm LJ}$ is immediately decreased and becomes almost constant until $t \sim 0.05 \times 7500 \tau$.
This step indicates the adsorption of the NP on the lipid membrane.
The sudden energy decrease by the NP adsorption is found in all three calculations as shown in Fig. S6~\cite{SI}.
Just after $t \sim 0.05 \times 7500 \tau$, the potential $V^{\rm (NL)}_{\rm LJ}$ is dramatically decreased again.
This process is the permeation of the NP through the lipid membrane.
This energy drop is also found in other two calculations at $t=0.02 \times 7500 \tau$ and $t=0.07 \times 7500 \tau$ (Fig. S6~\cite{SI}) which are almost the same time scale as $t=0.05 \times 7500 \tau$.
For NP passage through the bilayer membrane, lipids are pushed away by NP.
The displaced lipids return to fill the hole at $t=0.18 \times 7500 \tau$, in other words, the defect made by NP passage is completely fixed. 
The NP that has passed through the membrane is adsorbed on the inner leaflet.
The potential $V^{\rm (NL)}_{\rm LJ}$ gradually decreases from $t=0.18 \times 7500 \tau$ because the vesicle shape elongates to fit the NP (Fig.~\ref{fig1}(a)).
The electrostatic interaction between charged lipids, $V^{\rm (LL)}_{\rm el}$, is not changed significantly (Fig.~\ref{fig1}(d)).
Therefore, the lipid distribution in the membrane is not changed by the adsorption or the permeation of the NP.

We note that the time scale of permeation ($t \simeq 0.1 \times 7500 \tau$) is much shorter than that of membrane relaxation ($t \simeq 2 \times 7500 \tau$) as shown in Fig. S1 (b)~\cite{SI}.
From Fig. S1~\cite{SI}, the attractive energy $V^{\rm (LL)}_{\rm at}$ is decreased from $-1.31 \times 10^{5} k_{\rm B}T$ to $-1.33 \times 10^{5} k_{\rm B}T$ and the decrease is $2 \times 10^{3} k_{\rm B}T$.
The energy difference per lipid is $2 \times 10^{3} k_{\rm B}T / 5000 = 0.4 k_{\rm B}T$, which is smaller than $v_{\rm nh}$ and $v_{\rm nt}$.
Therefore, during the permeation process, the interaction between NP and lipid is predominant as compared to lipid-lipid interaction, and the permeation time scale is almost determined only by the NP-lipid interaction.

We focus on the process of the NP permeation through the lipid membrane.
At $t=0.07 \times 7500 \tau$, the NP passes through the lipid membrane.
In Fig.~\ref{fig1}(e), the density correlation functions of A- and B-lipids around the NP are indicated by red and blue lines, respectively.
When the distance from the NP is defined as $r_{\rm n}$, these lines start from $r_{\rm n}/\sigma=2.5$ located on the surface of NP and go up to 0.6 immediately.
The values of density correlation functions keep around 0.6.
Therefore, these lines change in the same way, indicating that the A- and B-lipids are uniformly distributed around the NP.
Figure~\ref{fig1}(f) shows the density correlation functions of hydrophilic a-beads (red line), hydrophobic b-beads located at the center of the lipid molecules (blue line), and hydrophobic c-beads located at the end of the lipid molecules (green line) around the NP.
There are more hydrophilic beads near the NP and we can see the peak value 0.7 at $r_{\rm n}/\sigma=3.5$.

Finally, the NP is adsorbed on the inner leaflet and induces the slight elongation of the vesicle.
The NP adsorbed to the inner leaflet tends to deform the lipid membrane and come into contact with the membrane of a larger area.
As a result, the vesicle is elongated.
This is caused by competition between adsorption energy and curvature energy to bend the lipid membrane.
The angle which determines the adsorption area is shown in Fig. S11~\cite{SI}, and it is $213.5 \pm 1.0^{\circ}$ for $z_{\rm n}=0, v_{\rm nh}/k_{\rm B}T=2, v_{\rm nt}/k_{\rm B}T=3$.
In addition, the density correlation functions of A- and B-lipids in the inner and outer leaflets around the NP are shown in Figs.~\ref{fig1}(g) and (h), respectively.
In the inner leaflet, there is a slight increase in neutral lipids (B-lipids) near the NP.
The inner leaflet near the NP has a large negative curvature to adsorb to the lipid membrane, and the charged headgroups are close to each other.
Consequently, it is difficult for charged lipids (A-lipids) to exist in this region due to the large electrostatic repulsion between charged lipids (A-lipids).
On the other hand, in the outer leaflet, there are some charged lipids (A-lipids) near the NP.
Because the outer leaflet has a positive curvature, the electrostatic repulsion is decreased, allowing charged lipids (A-lipids) to localize in this region.
However, the effect of lipid accumulation induced by the vesicle shape deformation is limited.

The NP is placed near the outer leaflet and spontaneously enters the interior of the vesicle.
Because the vesicle size is small, the NP on the inner leaflet can interact with many lipid beads without large membrane deformation; the curvature of NP fits the inner membrane easily.
The Lennard--Jones potential in the inner leaflet is smaller than that in the outer leaflet (Fig.~\ref{fig1}(b)), which is the driving force that allows the NP to enter the vesicle spontaneously.

\begin{figure}[t!]
\begin{center}
\includegraphics{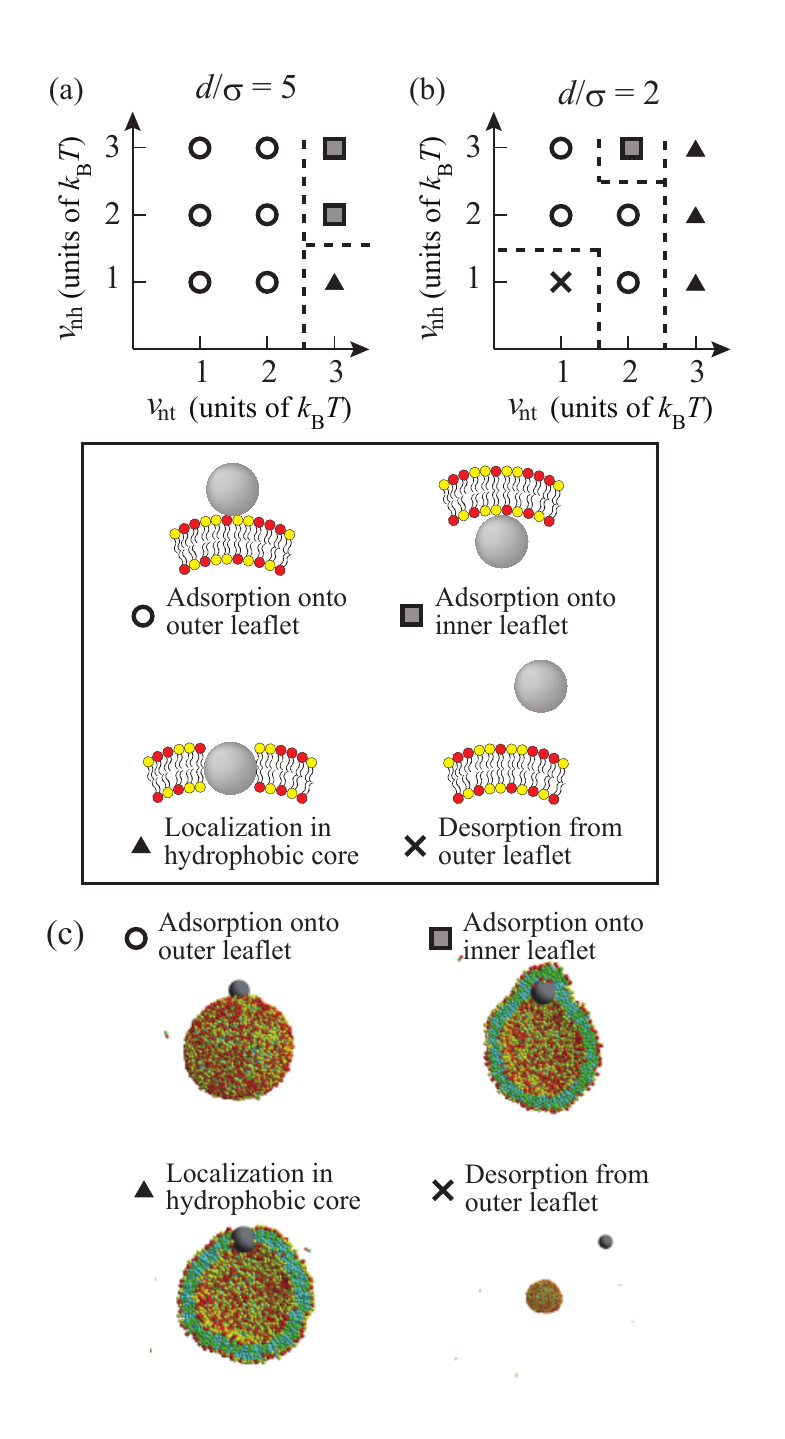}
\end{center}
\caption{
(a,b) Summary of neutral NP behavior at $t=1 \times 7500 \tau$ for $d/\sigma=5$ and $2$, respectively.
Open circles, gray squares, filled triangles, and the cross indicate adsorption on the outer leaflet, adsorption on the inner leaflet, localization in the hydrophobic core, and desorption from the outer leaflet, respectively.
Dashed lines roughly indicate the boundaries of the behaviors.
Schematic illustration of NP behavior is shown in a square box.
(c) Typical snapshots of NP behavior.
}
\label{fig2}
\end{figure}

We summarize the dynamic behavior of a neutral NP with $d/\sigma=5$ in Fig.~\ref{fig2}(a) and with $d/\sigma=2$ in Fig.~\ref{fig2}(b).
Typical snapshots of the final state at $t=1 \times 7500 \tau$ are shown in Fig.~\ref{fig2}(c).
When the interaction parameter between the NP with $d/\sigma=5$ and the hydrophobic beads, $v_{\rm nt}/k_{\rm B}T$, is less than $2$, the NP is adsorbed on the outer leaflet and does not enter the inside of the membrane or the vesicle (Fig.~\ref{fig2}(a)).
At $v_{\rm nt}/k_{\rm B}T=3$, the NP can disturb the membrane structure and can remain in or pass through the lipid membrane.
On the other hand, a smaller NP with $d/\sigma=2$ is not adsorbed on the outer leaflet when the interaction parameters are too small ($v_{\rm nh}/k_{\rm B}T=v_{\rm nt}/k_{\rm B}T=1$).
Because the small NP interacts with only a few lipid beads, it desorbs from the outer leaflet.
Unlike the larger NP, the smaller NP can adsorb on the inner leaflet for $v_{\rm nh}/k_{\rm B}T=3$ and $v_{\rm nt}/k_{\rm B}T=2$.
Moreover, the smaller NP tends to localize in the hydrophobic region of the lipid membrane as it does not disturb the membrane structure greatly.
Some snapshots and density correlation functions for $d/\sigma=2$ are shown in Fig. S3~\cite{SI}.

\subsection{Charged NP}
\label{charged}

\begin{figure*}[t!]
\begin{center}
\includegraphics{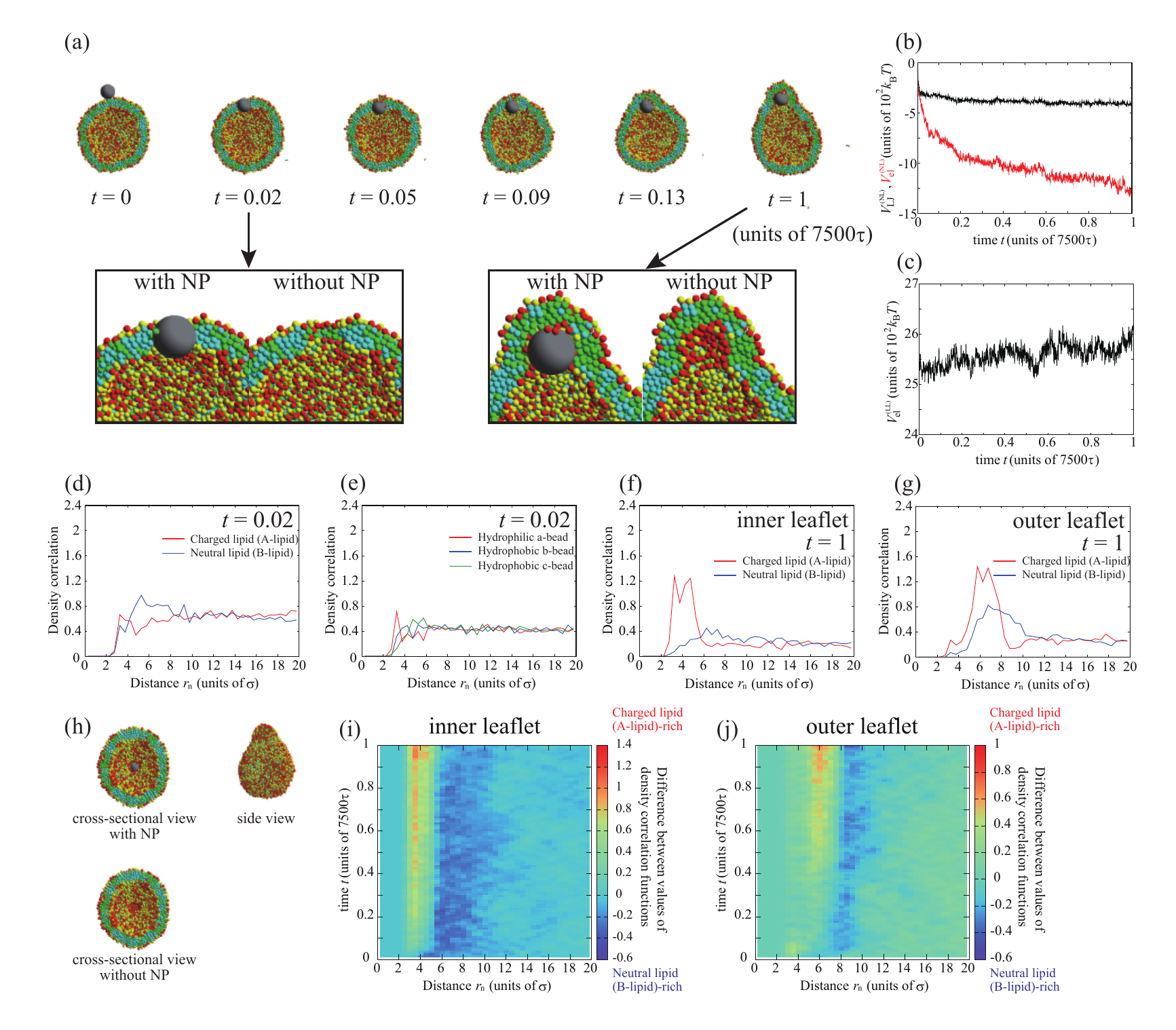}
\end{center}
\caption{
(a) Typical snapshots of the permeation sequence of a charged NP ($z_{\rm n}=+50$) for $d/\sigma=5$, $v_{\rm nh}/k_{\rm B}T=2$, and $v_{\rm nt}/k_{\rm B}T=3$.
Charged lipid (A-lipid) is represented by red hydrophilic beads and green hydrophobic beads, while neutral lipid (B-lipid) is represented by yellow hydrophilic beads and cyan hydrophobic beads.
(b) Time evolution of the Lennard--Jones potential between the NP and lipids, $V^{\rm (NL)}_{\rm LJ}$ (black line), and the electrostatic interaction between the NP and lipids, $V^{\rm (NL)}_{\rm el}$ (red line), for $t=1 \times 7500\tau$.
(c) Time evolution of the electrostatic interaction between lipids $V^{\rm (LL)}_{\rm el}$ for $t=1 \times 7500\tau$.
(d,e) Density correlation functions of lipids and beads around the NP at $t=0.02 \times 7500\tau$, respectively.
Red and blue lines in (d) indicate the density correlation functions of A- and B-lipids, respectively.
Red, blue, and green lines in (e) indicate the density correlation functions of a-, b-, and c-beads, respectively.
(f,g) Density correlation functions of A- and B-lipids in the inner leaflet in (f) and outer leaflet in (g) around the NP at $t=1 \times 7500\tau$.
Red and blue lines indicate the density correlation functions of A- and B-lipids, respectively.
(h) Snapshots at $t=1 \times 7500 \tau$.
(i,j) Lipid distribution changes in inner and outer leaflets as functions of distance and time, respectively. Color represents lipid distribution, and positive and negative values indicate charged lipid (A-lipid)-rich region and neutral lipid (B-lipid)-rich region, respectively.
}
\label{fig3}
\end{figure*}

Next, we place a charged NP ($z_{\rm n} \neq 0$) on the outer leaflet.
First, the valence of an NP is set to $z_{\rm n}=+50$.
Typical dynamics for $d/\sigma=5$, $v_{\rm nh}/k_{\rm B}T=2$, and $v_{\rm nt}/k_{\rm B}T=3$ are shown in Fig.~\ref{fig3}.
The parameters, except for the NP valence, are the same as for the calculation in Fig.~\ref{fig1}.
To ensure data reproducibility, the results obtained by the other two calculations are shown in Fig. S7~\cite{SI}.
As shown in the snapshots in Fig.~\ref{fig3}(a), the charged NP passes through the lipid membrane and is adsorbed on the inner leaflet.
This behavior is similar to that of a neutral NP (Fig.~\ref{fig1}).
However, the membrane permeation speed for a charged NP is faster than that for a neutral NP.
For the charged NP, the NP is adsorbed on the inner leaflet at $t=0.13 \times 7500 \tau$, whereas the NP adsorption on the inner leaflet is observed from $t=0.18 \times 7500 \tau$ for the neutral NP.
The Lennard--Jones potential between the NP and lipids, $V^{\rm (NL)}_{\rm LJ}$ (black line in Fig.~\ref{fig3}(b)), immediately decreases and the stepwise decrease observed in Fig.~\ref{fig1}(b) does not occur.
In contrast to the neutral NP, the electrostatic interaction between the charged NP and the lipids, $V^{\rm (NL)}_{\rm el}$ (red line in Fig.~\ref{fig3}(b)), makes a strong contribution and the energy is significantly decreased.
The fast membrane permeation is attributed to this electrostatic energy gain.

We show the density correlation functions of A- and B-lipids around the NP in Fig.~\ref{fig3}(d) to investigate the process of the permeation through the lipid membrane ($t=0.02 \times 7500 \tau$).
Since the values of density correlation functions for charged lipid (A-lipid) and neutral lipid (B-lipid) are about 0.7 and 0.5 at $r_{\rm n}/\sigma=3$, respectively, the number of negatively charged lipids (A-lipids) is larger near the positively charged NP.
This may be caused by the electrostatic attraction between the NP and charged lipids (A-lipids).
Figure~\ref{fig3}(e) shows the density correlation functions of a- to c-beads that make up the lipids around the NP.
There are slightly more hydrophilic beads near the NP (0.8 at $r_{\rm n}/\sigma=3$).
This effect is also small, because the indistinguishable profile is found in the neutral NP ($z_{\rm n}$=0) case as shown in Fig.~\ref{fig1}(f).
The electrostatic attraction between the charged NP and charged lipids increases the membrane permeation speed; however, the permeation process is similar to that for the neutral NP.

We observe a crucial difference in the lipid distribution in the membrane between the charged and neutral NPs.
There is a charged lipid (A-lipid)-rich region in the inner leaflet near the charged NP (enlarged view in Fig.~\ref{fig3}(a) and cross-sectional view in Fig.~\ref{fig3}(h)).
Even in the outer leaflet, the charged lipids are accumulated in the protruding part of the vesicle (snapshot from the side view in Fig.~\ref{fig3}(h)).
In Fig.~\ref{fig3}(c), the electrostatic interaction between charged lipids, $V^{\rm (LL)}_{\rm el}$, increases over time.
This result also implies that the lipid distribution becomes heterogeneous.
The lipid accumulation is also indicated by the density correlation functions of A- and B-lipids in the inner and outer leaflets (Figs.~\ref{fig3}(f) and (g)).
In both leaflets, the charged lipids (A-lipids) accumulate around the charged NP due to the electrostatic attraction between charged lipids and the charged NP.
Although the electrostatic repulsion between charged lipids is increased by the charged lipid accumulation, the system obtains a large energetic gain from the electrostatic attraction between the charged NP and charged lipids.
Because the energy gain shown in Fig.~\ref{fig3}(b) overcomes the energy loss shown in Fig.~\ref{fig3}(c), the charged lipids accumulate near the NP.
Moreover, the angle which determines the NP adsorption area is $273.8\pm13.3^{\circ}$ as shown in Fig. S11~\cite{SI}.
The angle becomes larger than that for neutral NP due to the contribution of the electrostatic interaction.

Neutral lipids (B-lipids) are abundant at $r_{\rm n}/\sigma=7$ in the inner leaflet (Fig.~\ref{fig3}(f)), whereas there are fewer charged lipids (A-lipids).
Moreover, there are more neutral lipids (B-lipids) than charged lipids (A-lipids) around $r_{\rm n}/\sigma=9$ in the outer leaflet (Fig.~\ref{fig3}(g)).
This region appears outside the peak of the charged lipids (A-lipids) because the charged lipids (A-lipids) accumulate near the NP and are depleted in the outer region.
In particular, the difference between the amounts of charged and neutral lipids is significant in the outer leaflet.
In the outer leaflet, the outside of the region where the NP is adsorbed to the membrane, that is, the part constricted by the protrusion of the vesicle, has a negative curvature, and it is difficult for charged lipids to localize in this region.
Thus, there is a region that is slightly neutral lipid-rich in the constricted part (side-view snapshot in Fig.~\ref{fig3}(h)).

Figures~\ref{fig3}(i) and (j) show the lipid distribution changes as functions of distance from the NP $r_{\rm n}$ and time.
The difference in color indicates the difference in the values of density correlation functions between charged lipid (A-lipid) and neutral lipid (B-lipid).
Therefore, positive and negative values indicate abundance in charged lipids (A-lipids) and neutral lipids (B-lipids), respectively.
In the inner leaflet, a charged lipid (A-lipid)-rich region is formed around $r_{\rm n}/\sigma=3$ to $4$ as shown in Fig.~\ref{fig3}(i).
As the color in a charged lipid (A-lipid)-rich region gets darker over time, we can see that more charged lipids (A-lipids) are accumulated.
In addition, it is found that the charged lipid (A-lipid) is particularly depleted and the neutral lipid (B-lipid)-rich region is formed around $r_{\rm n}/\sigma=6$ to $8$, which is outside the charged lipid (A-lipid)-rich region.
In the outer leaflet, a charged lipid (A-lipid)-rich region appears at $r_{\rm n}/\sigma=3$ to $4$ at $t<0.05 \times 7500\tau$ as shown in Fig.~\ref{fig3}(j).
It can be seen that this region formed in a short time is a charged lipid (A-lipid)-rich region induced by NP adsorption to the outer leaflet.
The region enriched in charged lipids (A-lipids) moves away from the NP with the lapse of time.
This indicates that the NP passes through the lipid membrane and moves to the inner leaflet.
Therefore, the region rich in charged lipids (A-lipids) at $r_{\rm n}/\sigma=6$ to $7$ can clearly be seen after $t \simeq 0.4 \times 7500\tau$.
Likewise, a neutral lipid (B-lipid)-rich region is formed along the outside of this charged lipid (A-lipid)-rich region.

\begin{figure}[t!]
\begin{center}
\includegraphics{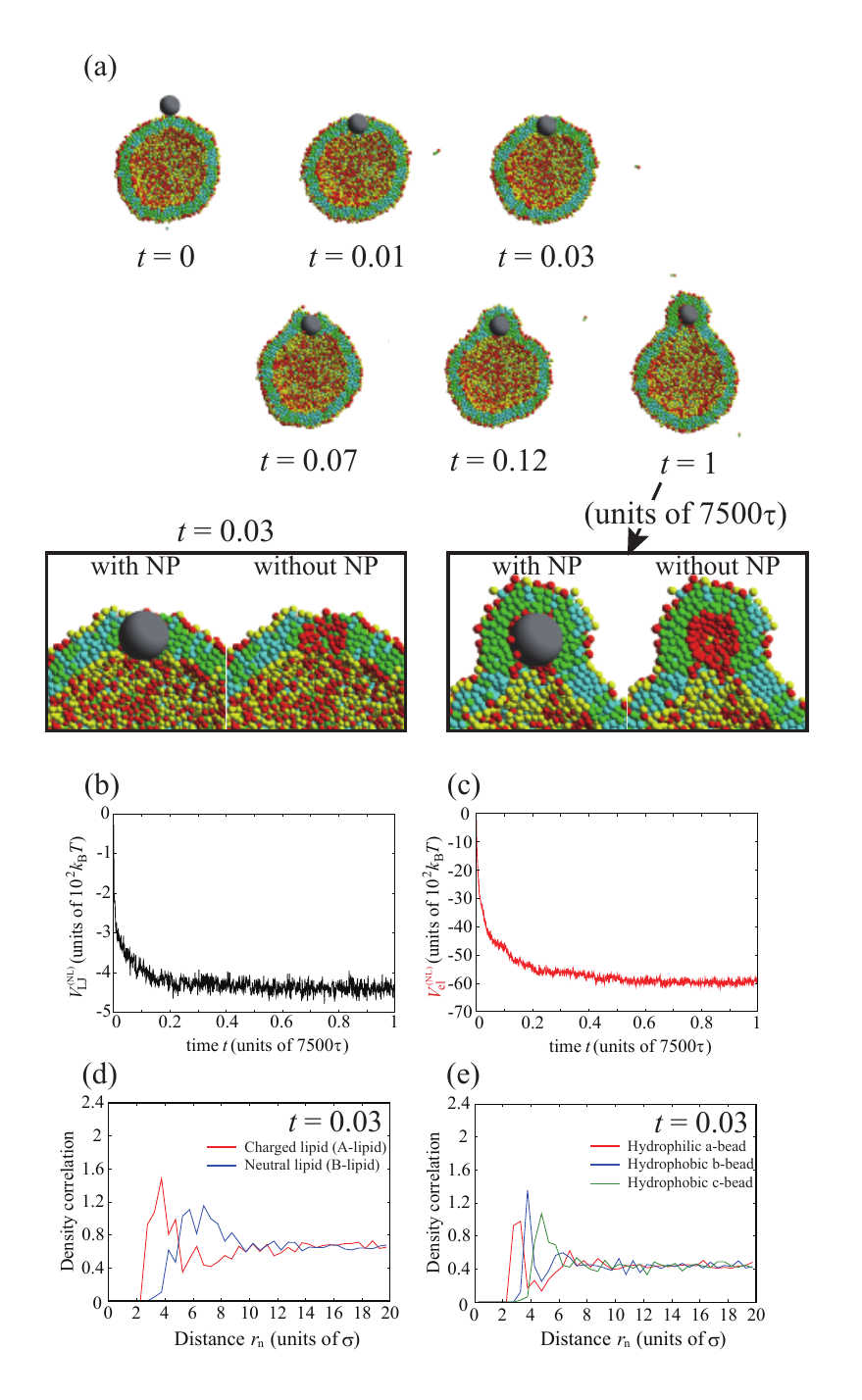}
\end{center}
\caption{
(a) Typical snapshots of the permeation sequence of a charged NP ($z_{\rm n}=+150$) through a transient channel for $d/\sigma=5$, $v_{\rm nh}/k_{\rm B}T=2$, and $v_{\rm nt}/k_{\rm B}T=3$.
Charged lipid (A-lipid) is represented by red hydrophilic beads and green hydrophobic beads, while neutral lipid (B-lipid) is represented by yellow hydrophilic beads and cyan hydrophobic beads.
(b) Time evolution of the Lennard--Jones potential between the NP and lipids, $V^{\rm (NL)}_{\rm LJ}$, for $t=1 \times 7500\tau$.
(c) Time evolution of the electrostatic interaction between the NP and lipids, $V^{\rm (NL)}_{\rm el}$, for $t=1 \times 7500\tau$.
(d,e) Density correlation functions of lipids and beads around the NP at $t=0.03 \times 7500\tau$, respectively.
Red and blue lines in (d) indicate the density correlation functions of the A- and B-lipids, respectively.
Red, blue, and green lines in (e) indicate the density correlation functions of the a-, b-, and c-beads, respectively.
}
\label{fig4}
\end{figure}

We consider an increase in the NP valence for the charged NP with $z_{\rm n}=+150$.
In Fig.~\ref{fig4}, we show the typical dynamic behavior of the charged NP with $z_{\rm n}=+150$, $d/\sigma=5$, $v_{\rm nh}/k_{\rm B}T=2$, and $v_{\rm nt}/k_{\rm B}T=3$.
The parameters, except for the valence of the NP, are the same as for the calculations in Figs.~\ref{fig1} and \ref{fig3}.
To ensure data reproducibility, the results obtained by the other two calculations are shown in Fig. S8~\cite{SI}.
The charged NP permeates through the lipid membrane and is adsorbed on the inner leaflet (Fig.~\ref{fig4}(a)).
This behavior is similar to the that of the neutral NP (Fig.~\ref{fig1}) and that of the charged NP with $z_{\rm n}=+50$ (Fig.~\ref{fig3}).
Because the valence of the NP ($z_{\rm n}=+150$) is larger than that in Fig.~\ref{fig3} ($z_{\rm n}=+50$), the permeation speed is also faster.
The fast permeation is mediated by the large electrostatic energy gain by the attraction between the charged NP and charged lipids (Fig.~\ref{fig4}(c)), although the change in the Lennard--Jones potential,  $V^{\rm (NL)}_{\rm LJ}$, in Fig.~\ref{fig4}(b) is similar to that shown in Fig.~\ref{fig3}(b).
In this case, there is also a charged lipid-rich region near the NP at $t=1 \times 7500 \tau$ as shown in the enlarged view of Fig.~\ref{fig4}(a).
In addition, the NP is completely wrapped by the lipid membrane at $t=1 \times 7500\tau$ as shown in Fig.~\ref{fig4}(a), and the adsorption angle becomes $360^{\circ}$ as shown in Fig. S11~\cite{SI}.

There is an interesting difference between the permeation processes for $z_{\rm n}=+50$ and $+150$.
The density correlation functions of A- and B-lipids near the NP at $t=0.03 \times 7500 \tau$ are shown in Fig.~\ref{fig4}(d) .
In the snapshot at $t=0.03 \times 7500 \tau$ (Fig.~\ref{fig4}(a)), the NP is just in the middle of the membrane.
In this step, the charged lipids (A-lipids) are accumulated near the charged NP (Fig.~\ref{fig4}(d)) and are depleted outside the peak.
Moreover, the beads closest to the NP are the hydrophilic a-beads, followed by the hydrophobic b-beads, and then the hydrophobic c-beads (Fig.~\ref{fig4}(e)), indicating that lipids are oriented with their hydrophilic heads toward the NP.
These results suggest that the charged lipids accumulate immediately near the NP, and the lipids are highly oriented toward the NP.
Therefore, a temporary channel covered with charged lipids is formed around the NP, the NP passes through the channel as shown in the enlarged view at $t=0.03 \times 7500\tau$ of Fig.~\ref{fig4}(a), 
and the channel disappears at $t=0.12 \times 7500 \tau$.

\begin{figure}[t!]
\begin{center}
\includegraphics{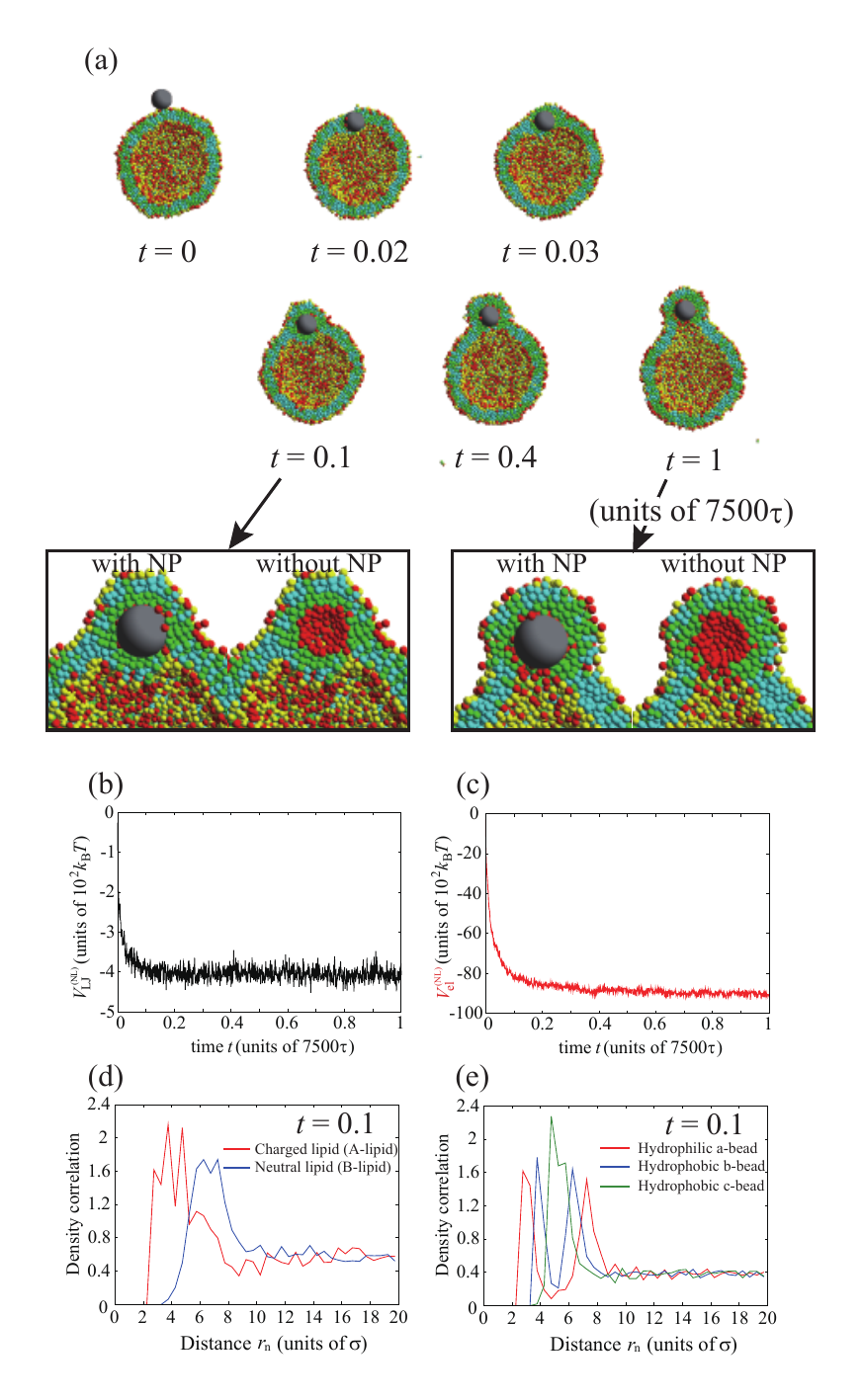}
\end{center}
\caption{(a) Typical snapshots of the permeation sequence of a charged NP ($z_{\rm n}=+200$) via transportlike endocytosis when $d/\sigma=5$, $v_{\rm nh}/k_{\rm B}T=2$, and $v_{\rm nt}/k_{\rm B}T=3$.
Charged lipid (A-lipid) is represented by red hydrophilic beads and green hydrophobic beads, while neutral lipid (B-lipid) is represented by yellow hydrophilic beads and cyan hydrophobic beads.
(b) Time evolution of the Lennard--Jones potential between the NP and lipids, $V^{\rm (NL)}_{\rm LJ}$, for $t=1 \times 7500\tau$.
(c) Time evolution of the electrostatic interaction between the NP and lipids, $V^{\rm (NL)}_{\rm el}$, for $t=1 \times 7500\tau$.
(d,e) Density correlation functions of lipids and beads around the NP at $t=0.1 \times 7500\tau$, respectively.
Red and blue lines in (d) indicate the density correlation functions of A- and B-lipids, respectively.
Red, blue, and green lines in (e) indicate the density correlation functions of a-, b-, and c-beads, respectively.}
\label{fig5}
\end{figure}

When the valence of a charged NP is increased to $z_{\rm n}=+200$, we observe a different permeation process through the lipid membrane.
The typical dynamic behavior of a charged NP with $z_{\rm n}=+200$, $d/\sigma=5$, $v_{\rm nh}/k_{\rm B}T=2$, and $v_{\rm nt}/k_{\rm B}T=3$ is shown in Fig.~\ref{fig5}.
The parameters, except for the valence of the NP, are the same for the calculations in Figs.~\ref{fig1}, \ref{fig3}, and \ref{fig4}.
To ensure data reproducibility, the results obtained by the other two calculations are shown in Fig. S9~\cite{SI}.
As shown in the snapshots in Fig.~\ref{fig5}(a), the NP is adsorbed on the outer leaflet immediately, buried in the membrane at $t=0.02 \times 7500 \tau$, and covered with the lipids at $t=0.03 \times 7500 \tau$.
At $t=0.1 \times 7500 \tau$, the NP is completely wrapped with the lipid membrane and the adsorption angle becomes $360^{\circ}$ as shown in Fig. S11~\cite{SI}.
After $t=0.1 \times 7500 \tau$, the vesicle shape is elongated to fit the NP, and the bypass bilayer, which is under the NP, disappears at $t=0.4 \times 7500 \tau$.
Finally, the NP is adsorbed on the inner leaflet of the membrane and the charged lipid-rich region is observed near the NP at $t=1 \times 7500 \tau$ as shown in Fig.~\ref{fig5}(a).
Because the NP is covered with the lipid membrane and is transported to the interior of the vesicle, we call this process transportlike endocytosis.

The Lennard--Jones potential between the NP and lipids, $V^{\rm (NL)}_{\rm LJ}$, in Fig.~\ref{fig5}(b) is similar to those in Figs.~\ref{fig3}(b) and \ref{fig4}(b).
There is a significant decrease in the electrostatic interaction between the NP and charged lipids, $V^{\rm (NL)}_{\rm el}$, in Fig.~\ref{fig5}(c).
Although the electrostatic interaction, $V^{\rm (NL)}_{\rm el}$, for $z_{\rm n}=+150$ also decreases dramatically, almost no transportlike endocytosis is observed.
This is because the energy gain by the electrostatic interaction, $V^{\rm (NL)}_{\rm el}$, does not compensate for the energy loss induced by the large membrane deformation for endocytic movement.

Here, we focus on the process of transportlike endocytosis.
From the density correlation functions of the A- and B-lipids near the NP at $t=0.1 \times 7500 \tau$ in Fig.~\ref{fig5}(d), the charged lipids (A-lipids) are accumulated near the NP by the strong electrostatic attraction.
This result is similar to that for the transient channel formation (Fig.~\ref{fig4}).
Figure~\ref{fig5}(e) shows the complicated peak appearance, with the hydrophilic a-beads beads closest to the NP, followed by the hydrophobic b-beads, hydrophobic c-beads, hydrophobic b-beads, and hydrophilic a-beads.
This peak pattern indicates bilayer formation around the NP, and the NP is covered with the lipid bilayer at $t=0.1 \times 7500 \tau$ (Fig. ~\ref{fig5}(a)).

\begin{figure*}[t!]
\begin{center}
\includegraphics{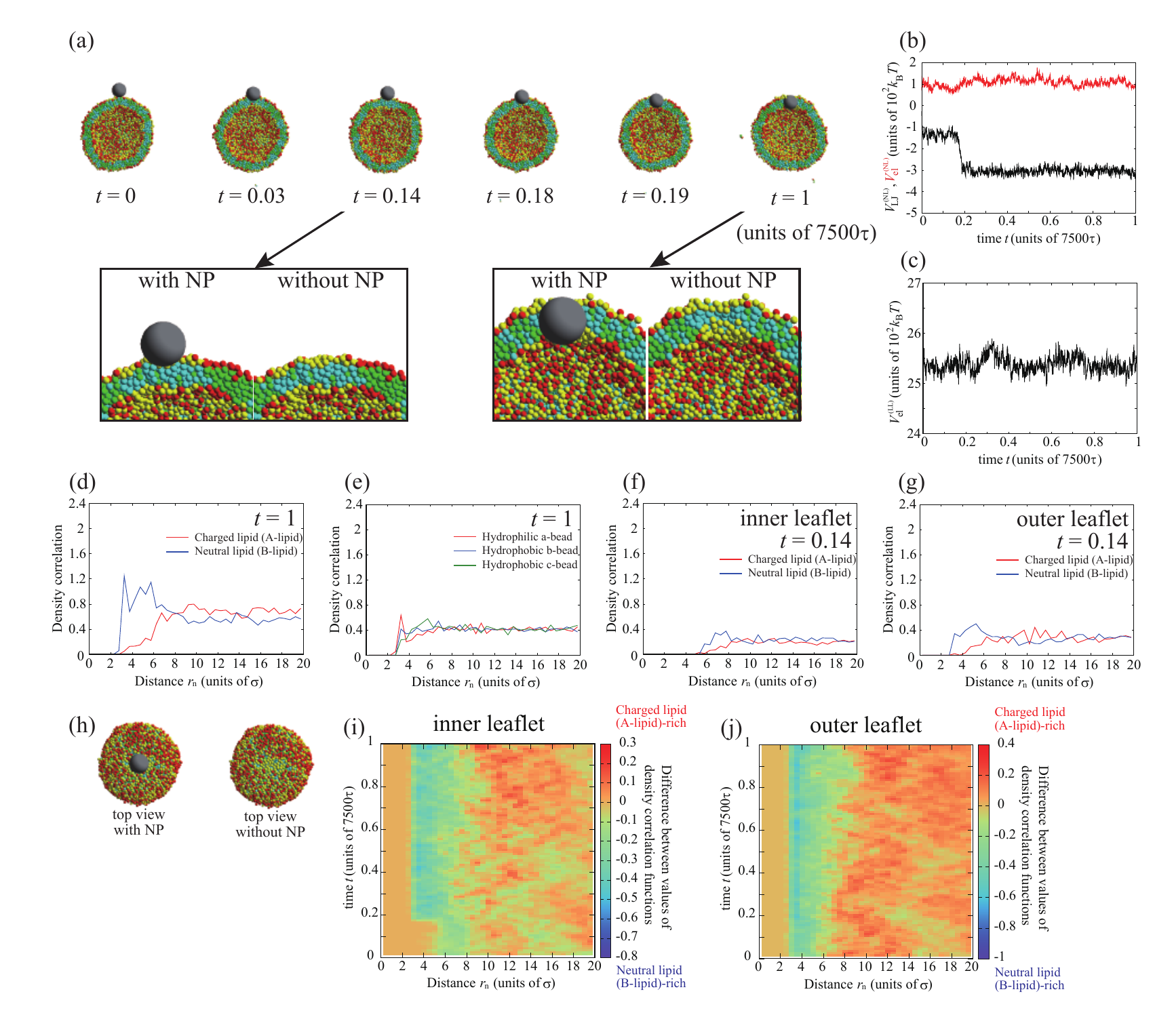}
\end{center}
\caption{(a) Typical snapshots of a negatively charged NP ($z_{\rm n}=-50$) when $d/\sigma=5$, $v_{\rm nh}/k_{\rm B}T=2$, and $v_{\rm nt}/k_{\rm B}T=3$.
Charged lipid (A-lipid) is represented by red hydrophilic beads and green hydrophobic beads, while neutral lipid (B-lipid) is represented by yellow hydrophilic beads and cyan hydrophobic beads.
(b) Time evolution of the Lennard--Jones potential between the NP and lipids, $V^{\rm (NL)}_{\rm LJ}$ (black line), and the electrostatic interaction between the NP and lipids, $V^{\rm (NL)}_{\rm el}$ (red line), for $t=1 \times 7500\tau$.
(c) Time evolution of the electrostatic interaction between lipids, $V^{\rm (LL)}_{\rm el}$, for $t=1 \times 7500\tau$.
(d,e) Density correlation functions of lipids and beads around the NP at $t=1 \times 7500\tau$, respectively.
Red and blue lines in (d) indicate the density correlations functions of A- and B-lipids, respectively.
Red, blue, and green lines in (e) indicate the density correlation functions of a-, b-, and c-beads, respectively.
(f,g) Density correlation functions of A- and B-lipids in the inner leaflet in (f) and outer leaflet in (g) around the NP at $t=0.14 \times 7500\tau$.
Red and blue lines indicate the density correlation functions of A- and B-lipids, respectively.
(h) Snapshots at $t=0.14 \times 7500 \tau$.
(i,j) Lipid distribution changes in inner and outer leaflets as functions of distance and time, respectively. Color represents lipid distribution, and positive and negative values indicate charged lipid (A-lipid)-rich region and neutral lipid (B-lipid)-rich region, respectively.}
\label{fig6}
\end{figure*}

Finally, we show the dynamic behavior of the charged NP with $z_{\rm n}=-50$, which has the same sign as the charged lipids.
Figure~\ref{fig6} shows the typical dynamic behavior of the charged NP with $z_{\rm n}=-50$, $d/\sigma=5$, $v_{\rm nh}/k_{\rm B}T=2$, and $v_{\rm nt}/k_{\rm B}T=3$.
The parameters, except for the valence of the NP, are the same as for the calculations in Figs.~\ref{fig1}, \ref{fig3}, \ref{fig4}, and \ref{fig5}.
To ensure data reproducibility, the results obtained by the other two calculations are shown in Fig. S10~\cite{SI}.
The NP is adsorbed on the outer leaflet and continues to adhere until $t \sim 0.14 \times 7500 \tau$ (Fig.~\ref{fig6}(a)).
After $t=0.14 \times 7500 \tau$, the NP starts to bury into the membranes and is localized in the hydrophobic region of the membrane at $t=0.19 \times 7500 \tau$ (Fig.~\ref{fig6}(a)).
The Lennard--Jones potential between the NP and the lipids, $V^{\rm (NL)}_{\rm LJ}$, and the electrostatic interaction between the NP and the lipids, $V^{\rm (NL)}_{\rm el}$, are shown in Fig.~\ref{fig6}(b).
The Lennard--Jones potential, $V^{\rm (NL)}_{\rm LJ}$, decreases quickly and the value remains almost constant until $t \sim 0.14 \times 7500 \tau$.
This step indicates the adsorption of the NP on the outer leaflet.
The Lennard--Jones potential decreases again from $t=0.14$ to $0.2 \times 7500 \tau$.
This process corresponds to the incorporation of the NP into the hydrophobic core of the membrane.
Because the charges of the NP and charged lipids (A-lipids) are the same sign, the electrostatic interaction, $V^{\rm (NL)}_{\rm el}$, has a positive value.
This repulsion tends to prevent the NP adsorption on the leaflets and the NP permeation through the membrane.

The density correlation functions at $t=1 \times 7500 \tau$ are shown in Fig.~\ref{fig6}(d) and (e).
There are many neutral lipids near the NP due to the electrostatic repulsion between the NP and the charged lipids (Fig.~\ref{fig6}(d)).
Although there is a slight increase in the number of hydrophilic beads near the NP (Fig.~\ref{fig6}(e)), the effect is small and the result is the same as that obtained for the neutral NP (Fig.~\ref{fig1}(f)).
Therefore, the NP is in the hydrophobic region of the lipid membrane and the orientation of the neutral lipids near the NP is not affected by the NP, and behaviors such as transient channel formation and transportlike endocytosis are not observed.

While the NP is adsorbed on the outer leaflet ($t=0.03 - 0.14 \times 7500 \tau$), a neutral lipid (B-lipid)-rich region forms just under the NP (Fig.~\ref{fig6}(h)).
This is because the charged lipids (A-lipids) near the NP are repelled from the region around the NP by the electrostatic repulsion and the neutral lipids (B-lipids) are present instead.
The concentration of charged lipids (A-lipids) in the outer region just under the NP hardly changes, even when the neutral lipid (B-lipid)-rich region is formed; thus, the electrostatic interaction between charged lipids (A-lipids), $V^{\rm (LL)}_{\rm el}$, in Fig.~\ref{fig6}(c) remains almost constant.
The density correlation functions in the inner and outer leaflets are shown in Figs.~\ref{fig6}(f) and (g).
In both leaflets, there are peaks for the neutral lipids near the NP.
Because the NP is adsorbed on the outer leaflet, the difference between the charged and neutral lipids is greater in the outer leaflet due to the large electrostatic repulsion.

We calculated the lipid distribution changes as functions of distance from the NP $r_{\rm n}$ and time as shown in Figs.~\ref{fig6}(i) and (j).
As in Figs.~\ref{fig3}(i) and (j), positive and negative values indicate charged lipid (A-lipid)-rich and neutral lipid (B-lipid)-rich regions, respectively.
Figure~\ref{fig6}(i) shows the lipid distribution change in the inner leaflet.
At $t<0.15 \times 7500\tau$, a region rich in neutral lipids (B-lipids) appears at $r_{\rm n}/\sigma=6$ to $8$.
This formation is caused by the adsorption of NP to the outer leaflet.
The region rich in neutral lipids (B-lipids) suddenly moves to $r_{\rm n}/\sigma=3$ to $5$ just before $t=0.2 \times 7500\tau$.
This indicates that the NP adsorbed on the membrane surface entered into the membrane.
From the snapshot of Fig.~\ref{fig6}(a), it can be seen that the NP localizes in the lipid membrane at $t=0.19 \times 7500\tau$.
Because the NP keeps the localization in the lipid membrane, the neutral lipid (B-lipid)-rich region does not migrate greatly after $t=0.2 \times 7500\tau$.
The lipid distribution change in the outer leaflet is shown in Fig.~\ref{fig6}(j).
A neutral lipid (B-lipid)-rich region is formed at $r_{\rm n}/\sigma=3$ to $5$, and this region does not migrate during the calculation.
The neutral lipid (B-lipid)-rich region is formed by the NP adsorption to the outer leaflet or the NP localization in the lipid membrane.
The distances between NP and the outer leaflet for these two situations are almost the same.
Therefore, it is considered that the position of the neutral lipid (B-lipid)-rich region is not changed.

\begin{figure*}[ht]
\begin{center}
\includegraphics{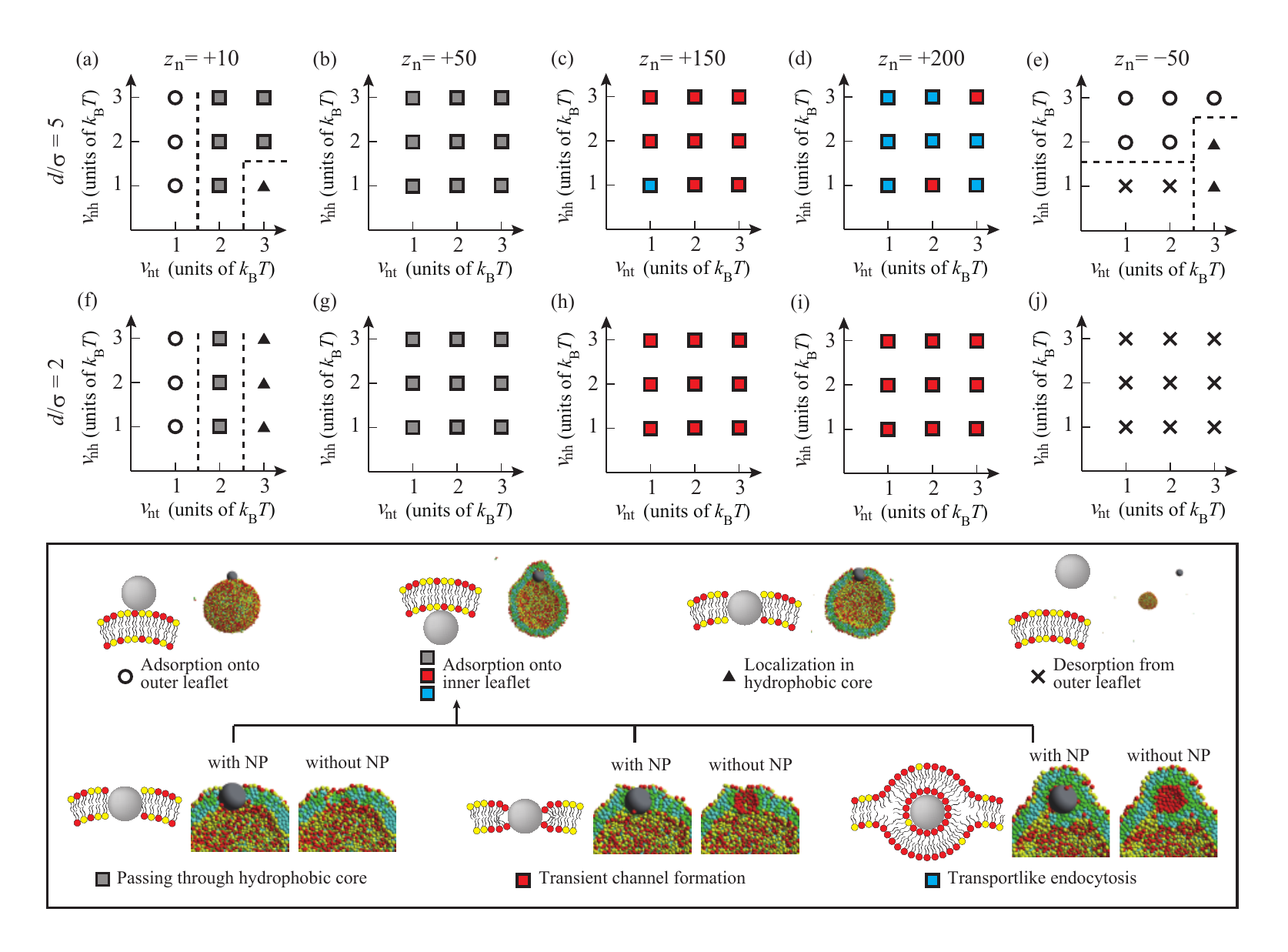}
\end{center}
\caption{
Summary of charged NP behavior at $t=1 \times 7500 \tau$ and uptake dynamics.
The valences and the diameters of NPs ($z_{\rm n}$, $d/\sigma$) are (a) (+10, 5), (b) (+50, 5), (c) (+150, 5), (d) (+200, 5), (e) (-50, 5), (f) (+10, 2), (g) (+50, 2), (h) (+150, 2), (i) (+200, 2), and (j) (-50, 2). 
Open circles, squares (gray, red, and blue), filled triangles, and crosses indicate adsorption onto the outer leaflet, adsorption onto the inner leaflet, localization in the hydrophobic core, and desorption from the outer leaflet, respectively.
Gray, red, and blue squares indicate permeation through the hydrophobic core, through the transient channel, and by transportlike endocytosis, respectively.
Dashed lines roughly indicate the boundaries of the behaviors.
Schematic and typical snapshot of the NP behaviors are shown at the bottom of the figure.}
\label{fig7}
\end{figure*}

We summarize the dynamic behavior of charged NPs with $d/\sigma=5$ and $2$ in Fig.~\ref{fig7}, and the valences and the diameters of the NPs ($z_{\rm n}$, $d/\sigma$) are (+10, 5) in (a), (+50, 5) in (b), (+150, 5) in (c), (+200, 5) in (d), (-50, 5) in (e), (+10, 2) in (f), (+50, 2) in (g), (+150, 2) in (h), (+200, 2) in (i), and (-50, 2) in (j). 
Some snapshots and density correlation functions for $d/\sigma=2$ are shown in Fig. S3~\cite{SI}.
For the neutral NP, roughly $v_{\rm nt}/k_{\rm B}T=3$ is needed to localize in the membrane or permeate through the membrane (Figs.~\ref{fig2}(a) and (b)).
The electrostatic attraction between the NP and lipids decreases the critical hydrophobic interaction strength for permeation through the membrane or localization in the membrane as $v_{\rm nt}/k_{\rm B}T=2$ (Figs.~\ref{fig7}(a) and (f)).
For $z_{\rm n}=+10$, no charged lipid (A-lipid)-rich region is formed just under the charged NP, because the electrostatic repulsion between the charged lipids when the lipids are accumulated is larger than the electrostatic attraction between the NP and lipids due to the low valence of the NP.
For $z_{\rm n}=+10$, the smaller NP with $d/\sigma=2$ is more likely to be present in the membrane, similar to the neutral NP.
This similarity arises because the small NP does not disturb the ordering of the lipids when it is present in the membrane.

When the valence of the NP is $z_{\rm n}=+50$, the NP passes through the hydrophobic region of the membrane and is adsorbed on the inner leaflet, regardless of the values of $v_{\rm nh}$ and $v_{\rm nt}$ (Figs.~\ref{fig7}(b) and (g)).
Although $z_{\rm n}=+50$ is too low to form a transient channel, it is high enough for the NP to pass through the membrane.
In Figs.~\ref{fig7}(c) and (h), permeation through the membrane occurs via the transient channel for $z_{\rm n}=+150$, except for $d/\sigma=5$ and $v_{\rm nh}/k_{\rm B}T=v_{\rm nt}/k_{\rm B}T=1$.
At $d/\sigma=5$ and $v_{\rm nh}/k_{\rm B}T=v_{\rm nt}/k_{\rm B}T=1$, transportlike endocytosis is the most common dynamic behavior.
When $z_{\rm n}$ is increased to $+200$, transportlike endocytosis occurs under many conditions for $d/\sigma=5$ (Fig.~\ref{fig7}(d)).
Interestingly, transportlike endocytosis is not observed for $z_{\rm n}=+200$ and $d/\sigma=2$ (Fig.~\ref{fig7}(i)).
This is because the lipid membrane must be curved sharply to cover the small NP with the lipid membrane in endocytosis, and the electrostatic attraction between the NP and lipids cannot compensate for the energy loss arising from this large deformation.

When $v_{\rm b}$ is small, especially $v_{\rm nh}$, it is difficult for the negatively charged NP with $z_{\rm n}=-50$ to be adsorbed on the outer leaflet by overcoming the electrostatic repulsion between the NP and charged lipids for $d/\sigma=5$ (Fig.~\ref{fig7}(e)).
If the Lennard--Jones potential is strong enough, the negatively charged NP can be adsorbed on the outer leaflet.
For large $v_{\rm nt}$, the NP can exist in the hydrophobic region of the lipid membrane through the favorable attraction between the NP and hydrophobic beads.
On the other hand, the smaller NP with $d/\sigma=2$ cannot be adsorbed on the outer leaflet (Fig.~\ref{fig7}(j)).
Because the number of lipid beads interacting with the NP is less for the smaller NP, a sufficient attractive interaction is not achieved and the NP desorbs from the outer leaflet.

\section{DISCUSSION}
\label{discussion}

Several points merit further discussion.
Some experimental studies have reported that cationic dendrimers are internalized into cells via endocytosis~\cite{Seib,Albertazzi}.
As the charge density of the dendrimers increased, the dendrimers were more easily internalized into cell~\cite{Albertazzi}.
This result is good agreement with our results (Fig.~\ref{fig7}), and endocytosis is frequently found for the large NP ($d/\sigma=5$) and high charge density NP ($z_{\rm n}=+200$) in our simulation.

In evaluating the relationship between experiments and our results, it is important to consider the behavior of multiple NPs and their cooperative motion.
Membrane protrusion was observed experimentally when NPs were added to vesicles ~\cite{Li}.
This behavior was explained by the adsorption of NPs on the outer leaflet.
However, we observed membrane protrusion induced by NP adsorption on the inner leaflet.
In our calculations, NPs are relatively large since the vesicle size is small.
Therefore, the vesicle may be just stretched to fit the NP.
Also in the experiments, it is difficult to identify the location of NPs by microscopy due to the optical resolution.
It is important to calculate the interaction between a larger vesicle and multiple NPs and directly compare experimental results and simulation results in near future.
Moreover, domain formation induced by protein adsorption has been observed experimentally~\cite{Hinderliter,McLaughlin,Scheve}.
Small domain formation has also been reported, and this is consistent with our findings and other simulation work~\cite{Yang_Li}.
Scheve {\it et al.} reported that numerous small domains are formed by steric repulsion between adsorbed proteins~\cite{Scheve}.
Calculations with multiple NPs may produce similar domain formation, which would be confirmed in future work.
Moreover, electrostatic repulsion as well as the steric repulsion between NPs could form positional ordering arrangements, such as two-dimensional crystals~\cite{Srivastava}.

Liu {\it et al.} investigated the interaction between charged NPs and vesicles consisting of zwitterionic lipids by CGMD simulations with a MARTINI force field~\cite{Liu}.
Although the net charge of the hydrophilic headgroup is neutral, they considered a headgroup with a dipole, such as phosphatidylcholine.
Charged NPs mainly interact with the positive charge on the headgroup, which is located near the membrane surface.
Consequently, the behavior of NPs depends on their charge density, $\rho$ ($e$ nm$^{-2}$); localization in the hydrophobic core is observed for $\rho=-1$ and $+1$, adsorption is observed for $\rho=-3.58$, adsorption or penetration is observed for $\rho=-7.51$, and desorption is observed for $\rho=+7.15$.
Based on the charge density of NPs, our NPs are $\rho=+0.26$ for ($z_{\rm n}$,$d/\sigma$) = (+10,5),  $\rho=+1.3$ for ($z_{\rm n}$,$d/\sigma$) = (+50,5), $\rho=+3.9$ for ($z_{\rm n}$,$d/\sigma$) = (+150,5),  $\rho=+5.2$ for ($z_{\rm n}$,$d/\sigma$) = (+200,5),  $\rho=-1.3$ for ($z_{\rm n}$,$d/\sigma$) = (-50,5),  $\rho=+1.6$ for ($z_{\rm n}$,$d/\sigma$) = (+10,2), $\rho=+8.1$ for ($z_{\rm n}$,$d/\sigma$) = (+50,2), $\rho=+24.4$ for ($z_{\rm n}$,$d/\sigma$) = (+150,2), $\rho=+32.5$ for ($z_{\rm n}$,$d/\sigma$) = (+200,2), and $\rho=-8.1$ for ($z_{\rm n}$,$d/\sigma$) = (-50,2).
These charge densities are summarized in Table S2~\cite{SI}. 
The behavior of the NPs summarized in Fig.~\ref{fig7} is mainly consistent with that reported by Liu {\it et al.}.
The essential differences between our study and Liu {\it et al.} are that we consider charged membranes (net charge of membrane is not zero), a more coarse-grained lipid molecule,
and 5000 lipid molecules, compared with 1182 used by Liu {\it et al.}.
Although we consider a simple lipid molecule consisting of three beads connected linearly by a spring, this model is sufficient to consider the behavior of NPs with a lipid membrane qualitatively.
Our model reproduces the dynamic behavior of NPs with a charged lipid membrane for a wide range of interaction parameters, and NP valences and sizes.

An NP on an outer leaflet spontaneously passes through the lipid membrane and is adsorbed on the inner leaflet (Fig.~\ref{fig1}) because the attractive energy between the NP and lipids is lower when the NP is adsorbed on the inner leaflet (Fig.~\ref{fig1}(b)).
This difference in attractive energy between the NP on the outer and inner leaflets arises from the lipid membrane curvature.
To evaluate the effect of the membrane curvature on the NP uptake behavior, we perform the same calculations using a smaller vesicle consisting of 1000 lipid molecules. 
The vesicle consists of 500 charged lipids (A-lipids) and 500 neutral lipids (B-lipids) and the symmetric monovalent salt concentration is 100 mM.
We also perform the calculation until $t = 10 \times 7500\tau$ to relax the lipid membrane structure sufficiently with $w_{\rm c}^{\rm AA}/\sigma=1.8$, $w_{\rm c}^{\rm BB}/\sigma=1.7$, and $w_{\rm c}^{\rm AB}/\sigma=1.575$.
The NP behavior is summarized in Fig. S2~\cite{SI}.
Some snapshots and density correlation functions are shown in Figs. S4 and S5~\cite{SI}.
The neutral NP and charged NPs with $z_{\rm n}=+10$ and $-50$ pass more easily through the membrane of the smaller vesicle (Figs.~\ref{fig2}, \ref{fig7}, and S2).
For $z_{\rm n}=+50$ and $+150$, there is no large difference in NP behavior between the larger and smaller vesicles because the electrostatic interaction is dominant and the effect of the membrane curvature is negligible.
However, the transportlike endocytosis, which is seen for $z_{\rm n}=+200$, is not observed for the smaller vesicle.
Because of the large membrane curvature, a large membrane deformation is needed to cover the NP surface for the smaller vesicle.
Because this deformation costs a large amount of energy, transportlike endocytosis is suppressed.

In this study, the Debye-H\"{u}ckel approximation is used as the electrostatic interaction, and it will be appropriate for low electric potential and/or high salt concentration.
For NPs with high valence, it is known that such a mean-field theory does not treat ion-ion correlation correctly.
Therefore, the calculation including the presence of ions explicitly will be important for future work.
Moreover, there is a hydrophobic core in the center of lipid membrane.
Although the electrostatic interaction across the lipid membrane is considered as one of the important interactions~\cite{Shimokawa2,Baciu,Wagner},
it is almost decoupled due to low dielectric constant of the hydrophobic part.
Harries and his coworkers proposed a theoretical model for describing the endocytosis of charged globular protein~\cite{Harries}.
In this model, they considered that the outer leaflet of the lipid bilayer membrane wrapping an NP does not electrostatically interact with the NP.
Since we did not consider the low dielectric nature of hydrophobic core, our simulation overestimates the electrostatic interaction.
However, our CGMD simulation qualitatively consistent with the theoretical model, where the endocytosis occurs for charged protein with higher valence~\cite{Harries}.
For further quantitative comparison of our results with experimental and theoretical studies, we should propose a model including the ion-ion correlation and the low dielectric nature of hydrophobic core in future.

In the present study, we examine the vesicle configuration and the effect of the membrane curvature is different from that of the flat membrane.
For example, NPs are adsorbed onto inner leaflet under various conditions (Figs.~\ref{fig1}, \ref{fig3}, \ref{fig4}, and \ref{fig5}) because the curvature of NPs and that of inner leaflet fit, and thus the NPs can interact with more lipid beads in the inner leaflet without large membrane deformation.
Therefore, this uptake mechanism is mediated by the membrane curvature.
Since the vesicular diameter is on the order of 10 nm and it is much smaller than the cell size, which is on the micrometer scale. It is difficult to directly extrapolate the obtained results to the situation of much larger vesicles or cells.
However, the cell is not completely spherical, and has various local curvatures.
NP uptake mediated by membrane curvature could occur at the curved parts in the cell.
Moreover, Harries et al. indicated that the inner leaflet composition of the lipid membrane wrapping an NP should be satisfied with the condition of charge matching with NP~\cite{Harries}.
In our calculation, the number of lipids which can contact with NP is about 100 for $d/\sigma=5$.
Although the condition of charge matching is $z_{\rm n} \simeq +100$, we can see the endocytosis from $z_{\rm n}=+200$.
We consider two effects to explain this difference.
First, the theoretical model does not take into account the intermediate state between spherical protein adsorbing state and protein wrapped state.
Since the intermediate state is generally in a very high energy state, it cannot be discussed just by comparing the energies of the two stable states.
Second, it is considered that NPs in our calculations are too small to collect the sufficient number of charged lipids on the NP surface, 
and it can be difficult to gain sufficient electrostatic energy that overcomes the deformation energy for endocytosis.
Based on these points, it is also important to check the behaviors of larger vesicles, which needs to increase the speed and scale of calculation.
In future work, we will perform similar calculations in vesicles of different sizes to reveal more detailed dependence of the NP uptake behavior on the membrane curvature as well as endocytosis.

\section{CONCLUSION}
\label{conclusion}

In this paper, we report the interaction between charged NPs and negatively charged binary lipid vesicles based on a CGMD simulation.
For a neutral NP, the NP can be incorporated into the membrane hydrophobic region and the interior of a vesicle when the hydrophobic interaction parameter, $v_{\rm nt}$, is large enough.
A positively charged NP can permeate through the membrane owing to the electrostatic interaction, even at smaller $v_{\rm nt}$.
Moreover, at higher valences, membrane permeation can occur easily and the NP is transported via a transient charged lipid channel or endocytosis.
We can classify the permeation processes as passing through the hydrophobic core, transient channel formation, and transportlike endocytosis based on the density correlation function.
When the NP is small ($d/\sigma=2$), transportlike endocytosis does not occur because a large membrane deformation is necessary.
A negatively charged NP can be adsorbed onto the membrane for the large hydrophilic parameter, $v_{\rm nh}$, region.
The spontaneous uptake into the vesicle is caused by membrane curvature.
Using the CGMD model, we investigated the NP behavior with a combination of various parameters.
Another interesting phenomenon is that the NP delivered into a cell via endocytosis is released from the endosome~\cite{Vacha2}.
Based on our model, the release of NP due to the change in the electrostatic interaction can be examined by considering the difference in ionic strengths between inside and outside of the cell.
In future, the cooperative behavior of multiple NPs, the NP uptake behavior for a phase-separated membrane, and the NP uptake in larger and smaller vesicles will be investigated to discuss the mechanism of NP uptake into cells and vesicles in more detail.


\begin{acknowledgments}
Most of the calculations were performed using the parallel computer ``SGI UV3000'' provided by the Center for Information Science at JAIST. 
We acknowledge support from the Bilateral Joint Research Project (Japan-Slovenia) from the Japan Society for the Promotion of Science (JSPS). 
N.S. acknowledges support from a Grant-in-Aid for Scientific Research (C) 
(Grant No. JP17K05610) from JSPS. 
H.I. acknowledges support from a Grant-in-Aid for JSPS Research Fellow (Grant No. JP18J00259). 
\end{acknowledgments}

\newpage

\newpage

\clearpage
\appendix
\setcounter{figure}{0}
\renewcommand{\thefigure}{S\arabic{figure}}
\onecolumngrid
\section*{Supporting information}

\subsection*{Parameters of calculations}

\begin{table}[h]
\begin{center}
\caption{Calculation parameters of all figures in main text.}
\begin{tabular}{ccccc} \toprule
~Figure~ & ~$z_{\rm n}$~ & ~$d/\sigma$~ & ~$v_{\rm nh}/k_{\rm B}T$~ & ~$v_{\rm nt}/k_{\rm B}T$~ \\ \midrule
Fig. 1 & 0 & 5 & 2 & 3 \\
Fig. 2 & 0 & 5, 2 & 1--3 & 1--3 \\
Fig. 3 & +50 & 5 & 2 & 3 \\
Fig. 4 & +150 & 5 & 2 & 3 \\
Fig. 5 & +200 & 5 & 2 & 3 \\
Fig. 6 & -50 & 5 & 2 & 3 \\
Fig. 7 & \begin{tabular}{c}+10, +50, +150,\\ +200, -50 \end{tabular} & 5,2 & 1--3 & 1--3 \\ \bottomrule
\end{tabular}
\end{center}
\end{table}

\subsection*{Surface charge densities of NPs}

\begin{table}[h]
\begin{center}
\caption{Surface charge density of charged NPs.}
\begin{tabular}{ccc} \toprule
 ~$z_{\rm n}$~ & ~$d/\sigma$~ & ~$\rho$ ($e$ nm$^{-2}$)~ \\ \midrule
+10 & 5 & +0.26 \\
+50 & 5 & +1.3 \\
+150 & 5 & +3.9 \\
+200 & 5 & +5.2 \\
-50 & 5 & -1.3 \\
+10 & 2 & +1.6 \\
+50 & 2 & +8.1 \\
+150 & 2 & +24.4 \\
+200 & 2 & +32.5 \\
-50 & 2 & -8.1 \\ \bottomrule
\end{tabular}
\end{center}
\end{table}

\newpage

\subsection*{Relaxation of membrane structure}

Figure S1 (a) shows snapshots of the membrane structure in the binary 
charged lipid mixture composed of 2500 charged lipids (A-lipids) (red or dark) and 
2500 neutral lipids (B-lipids) (yellow or light), where the cutoff lengths for the 
attractive interactions among hydrophobic beads were set to 
$w_{\rm c}^{\rm AA}/\sigma=1.8$, $w_{\rm c}^{\rm BB}/\sigma=1.7$, and
$w_{\rm c}^{\rm AB}/\sigma=1.55$.
The membrane separates into A-lipid-rich and B-lipid-rich phases at 
$t=1 \times 7500 \tau$, and there is a clearer phase-separated structure 
at $t=10 \times 7500 \tau$. 
The change in the attractive potential, $V_{\rm{at}}^{\rm{(LL)}}$, is shown
 in Fig. S1(b). 
As the phase separation progresses, the potential decreases substantially 
until  $t \simeq 2 \times 7500 \tau$. 
At this time, the vesicle may reach the equilibrium state because the 
potential does not change significantly after $t \simeq 2 \times 7500 \tau$. 
Figure S1(c) shows the density correlation functions between lipid species 
at $t=10 \times 7500 \tau$. 
The density autocorrelation function of A-lipids (red line) intersects with 
the density cross-correlation function between the A- and B-lipids (green line). 
The spatial distance at this intersection is the typical domain size, and this 
result also indicates that phase separation occurs.
When the cutoff length, $w_{\rm c}^{\rm AB}/\sigma$, is changed to 
$w_{\rm c}^{\rm AB}/\sigma=1.575$, no phase separation occurs (Fig. S1(d)). 
No decrease in the attractive potential, $V_{\rm{at}}^{\rm{(LL)}}$, is visible 
immediately after the start of the calculation in Fig. S1(e), and the density 
autocorrelation function of the A-lipids (red line) and the density 
cross-correlation function between the A- and B-lipids (green line) do not 
intersect at $t=10 \times 7500 \tau$ in Fig. S1(f). 
Therefore, the phase separation does not occur for 
$w_{\rm c}^{\rm AB}/\sigma=1.575$. 
$V_{\rm{at}}^{\rm{(LL)}}$ continuously increases during the calculation in 
Fig. S1(e) because the lipids gradually escape from the vesicle. 
We regard the structure at $t=10 \times 7500 \tau$ as the static states of 
the preassembled vesicles; such states observed experimentally have been 
treated as practically static because of the much slower time scale of the 
complete melting of the structure. 
Our results are essentially the same as those reported in a previous study~\cite{Ito}.

\begin{figure}[h!]
\begin{center}
\includegraphics{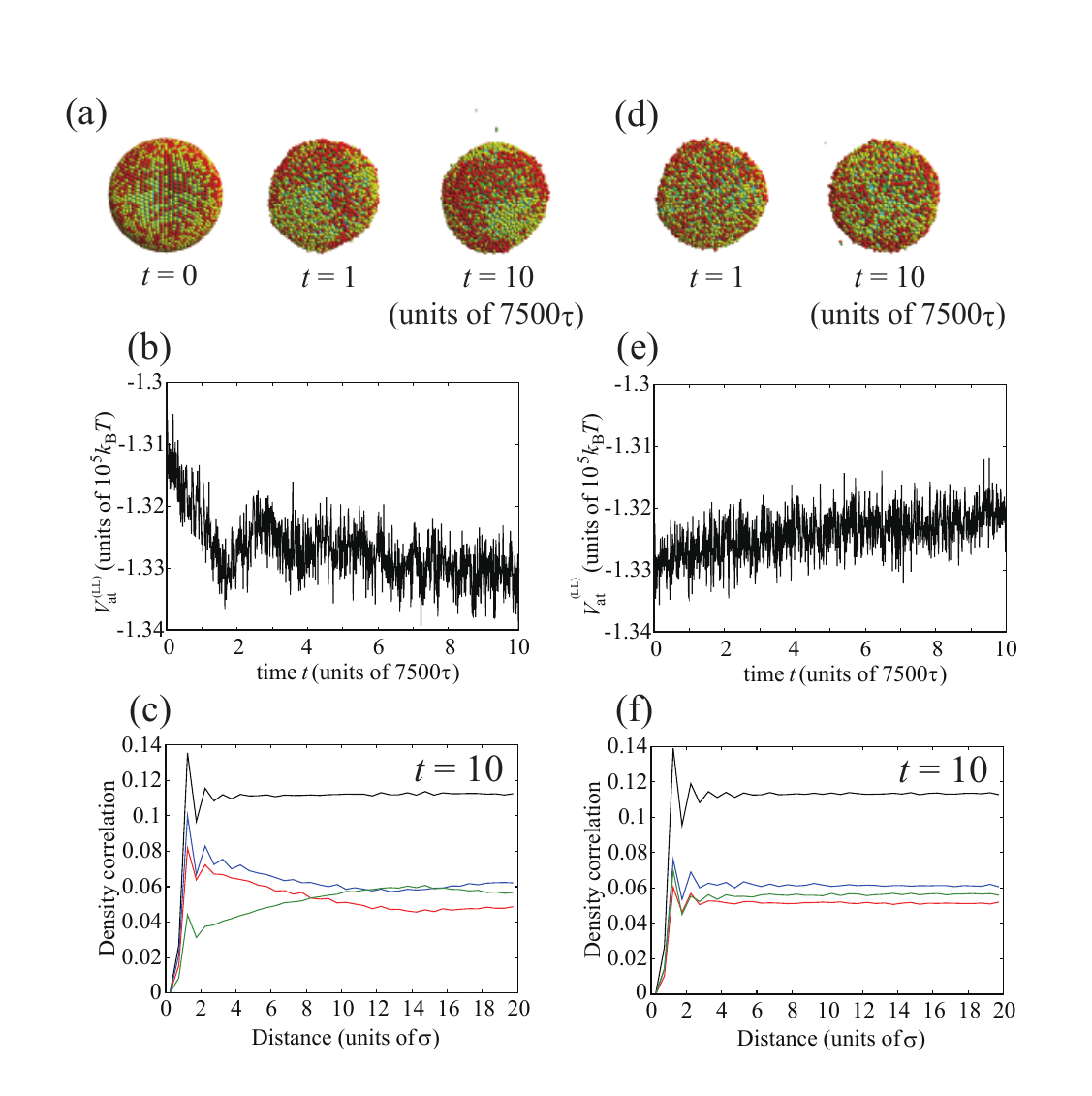}
\end{center}
\caption{
(a,d) Typical snapshots of a charged vesicle at initial state ($t=0$), $t=1 \times 7500 \tau$, and 
$t=10 \times 7500 \tau$ for $w_{\rm c}^{\rm AB}/\sigma=1.55$ in (a) and 
$w_{\rm c}^{\rm AB}/\sigma=1.575$ in (d). 
Charged lipid (A-lipid) is represented by red hydrophilic beads and green hydrophobic beads, while neutral lipid (B-lipid) is represented by yellow hydrophilic beads and cyan hydrophobic beads.
(b,e) Time evolution of the attractive potential $V_{\rm{at}}^{\rm{(LL)}}$ for 
$t=10 \times 7500 \tau$ for $w_{\rm c}^{\rm AB}/\sigma=1.55$ in (b) and 
$w_{\rm c}^{\rm AB}/\sigma=1.575$ in (e). 
(c,f) Density correlation function between lipid species at $t=10 \times 7500 \tau$ 
for $w_{\rm c}^{\rm AB}/\sigma=1.55$ in (c) and $w_{\rm c}^{\rm AB}/\sigma=1.575$ in (f). 
Black, red, and blue lines indicate the density autocorrelation functions of all lipids, 
A-lipids, and B-lipids, respectively. 
Green line indicates the density cross-correlation function between A- and B-lipids.}
\end{figure}

\newpage
\clearpage

\subsection*{Calculation for smaller vesicle}

\begin{figure}[h!]
\begin{center}
\includegraphics{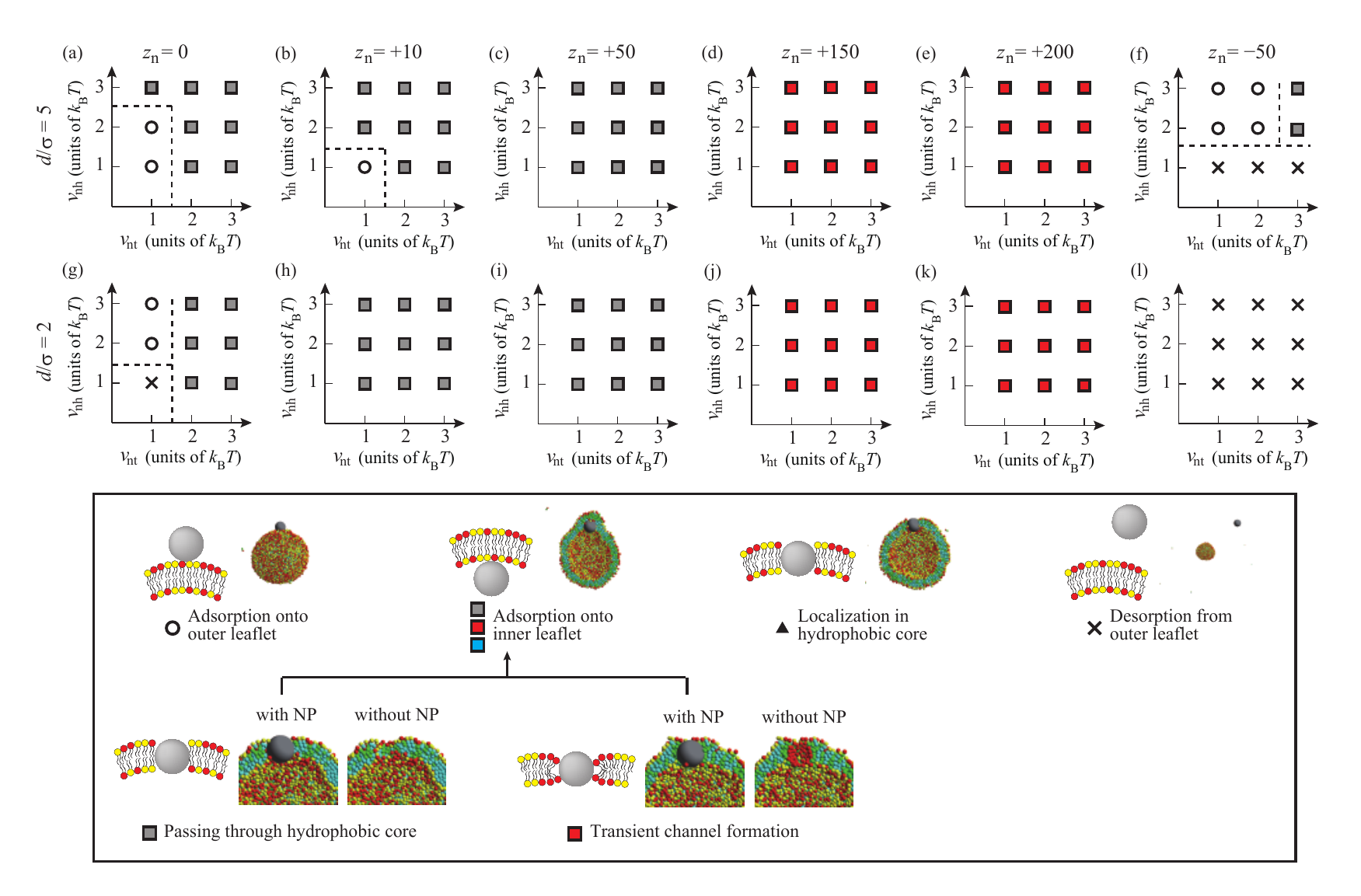}
\end{center}
\caption{
Summary of charged NP behavior at $t=1 \times 7500 \tau$ and uptake dynamics 
for a smaller vesicle consisting of 500 charged lipids (A-lipids) and 500 neutral lipids (B-lipids). 
The valences and the diameters of NPs ($z_{\rm{n}}$, $d/\sigma$) are (a) (0, 5), 
(b) (+10, 5), (c) (+50, 5), (d) (+150, 5), (e) (+200, 5), (f) ($-$50, 5), (g) (0, 2), (h) (+10, 2), 
(i) (+50, 2), (j) (+150, 2), (k) (+200, 2), and (l) ($-$50, 2). 
Open circles, squares (gray and red), filled triangles, and crosses indicate adsorption 
onto the outer leaflet, adsorption onto the inner leaflet, localization in the hydrophobic 
core, and desorption from the outer leaflet, respectively. 
Gray and red squares indicate the permeation through the hydrophobic core and 
through the transient channel, respectively. 
Dashed lines roughly indicate the boundaries of the behaviors. 
A schematic of the NP behavior is shown in the square box.}
\end{figure}

\newpage
\subsection*{Snapshots and density correlation functions}
\begin{figure}[h!]
\begin{center}
\includegraphics{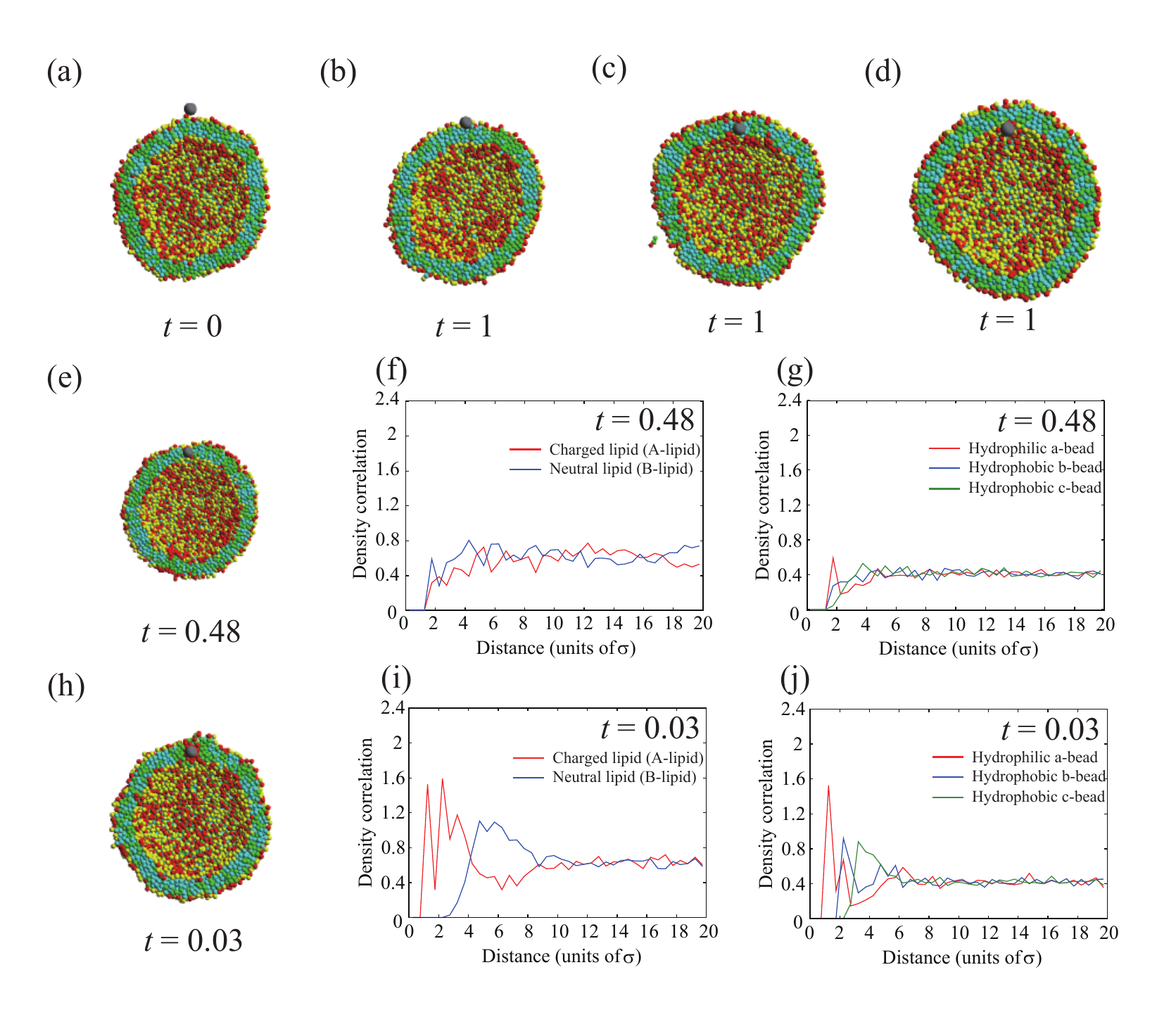}
\end{center}
\caption{
(a) Snapshot of initial state ($t=0 \times 7500 \tau$) for a vesicle composed of 2500 
charged lipids (A-lipids) and 2500 neutral lipids (B-lipids) with an NP ($d/\sigma = 2$). 
Charged lipid (A-lipid) is represented by red hydrophilic beads and green hydrophobic beads, while neutral lipid (B-lipid) is represented by yellow hydrophilic beads and cyan hydrophobic beads.
[(b), (c), and (d)] Typical snapshots at $t=1 \times 7500 \tau$.
(b) Adsorption onto outer leaflet, (c) localization in hydrophobic core, and (d) adsorption 
onto inner leaflet. 
(e) Typical snapshot when a neutral NP ($z_{\rm{n}} = 0$) passes through the hydrophobic 
core at $t=0.48 \times 7500 \tau$ for $d/\sigma = 2$, $v_{\rm nh}/k_{\rm B}T = 3$, and 
$v_{\rm nt}/k_{\rm B}T = 2$. 
[(f) and (g)] Density correlation functions of lipids and beads around the NP at $t=0.48 \times 7500 \tau$
corresponding to (e), respectively. 
Red and blue lines in (f) indicate the density correlation functions of A- and B-lipids, respectively. 
Red, blue, and green lines in (g) indicate the density correlation functions of a-, b-, and c-beads, 
respectively. 
(h) Typical snapshot when a positively charged NP ($z_{\rm {n}} = +150$) passes through a transient 
channel at $t=0.03 \times 7500 \tau$ for $d/\sigma = 2$, $v_{\rm nh}/k_{\rm B}T = 1$, and 
$v_{\rm nt}/k_{\rm B}T = 1$. 
[(i) and (j)] Density correlation functions of lipids and beads around the bead at $t=0.03 \times 7500 \tau$
corresponding to (h), respectively.}
\end{figure}

\clearpage

\begin{figure}[h!]
\begin{center}
\includegraphics{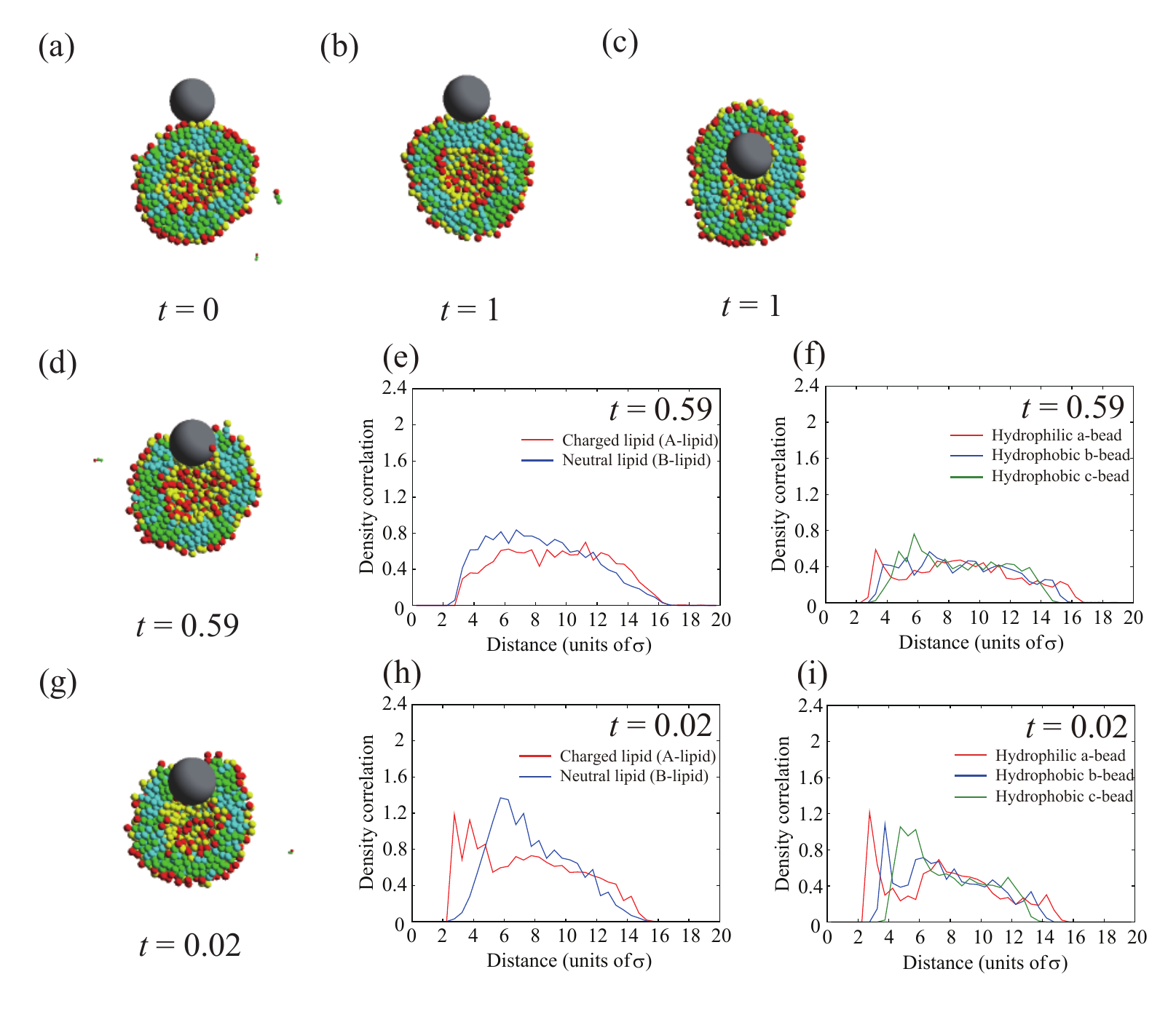}
\end{center}
\caption{
(a) Snapshot of initial state ($t=0 \times 7500 \tau$) for a vesicle composed of 500 
charged lipids (A-lipids) and 500 neutral lipids (B-lipids) with an NP ($d/\sigma = 5$). 
Charged lipid (A-lipid) is represented by red hydrophilic beads and green hydrophobic beads, while neutral lipid (B-lipid) is represented by yellow hydrophilic beads and cyan hydrophobic beads.
[(b) and (c)] Typical snapshots at $t=1 \times 7500 \tau$.
(b) Adsorption onto outer leaflet and (c) adsorption onto inner leaflet. 
(d) Typical snapshot when a neutral NP ($z_{\rm{n}} = 0$) passes through the hydrophobic 
core at $t=0.59 \times 7500 \tau$ for $d/\sigma = 5$, $v_{\rm nh}/k_{\rm B}T = 1$, and 
$v_{\rm nt}/k_{\rm B}T = 2$. 
[(e) and (f)] Density correlation functions of lipids and beads around the NP at $t=0.59 \times 7500 \tau$
corresponding to (d), respectively. 
Red and blue lines in (e) indicate the density correlation functions of A- and B-lipids, respectively. 
Red, blue, and green lines in (f) indicate the density correlation functions of a-, b-, and c-beads, 
respectively. 
(g) Typical snapshot when a positively charged NP ($z_{\rm {n}} = +150$) passes through a transient 
channel at $t=0.02 \times 7500 \tau$ for $d/\sigma = 5$, $v_{\rm nh}/k_{\rm B}T = 1$, and 
$v_{\rm nt}/k_{\rm B}T = 1$. 
[(h) and (i)] Density correlation functions of lipids and beads around the bead at $t=0.02 \times 7500 \tau$
corresponding to (g), respectively.}
\end{figure}

\clearpage

\begin{figure}[h!]
\begin{center}
\includegraphics{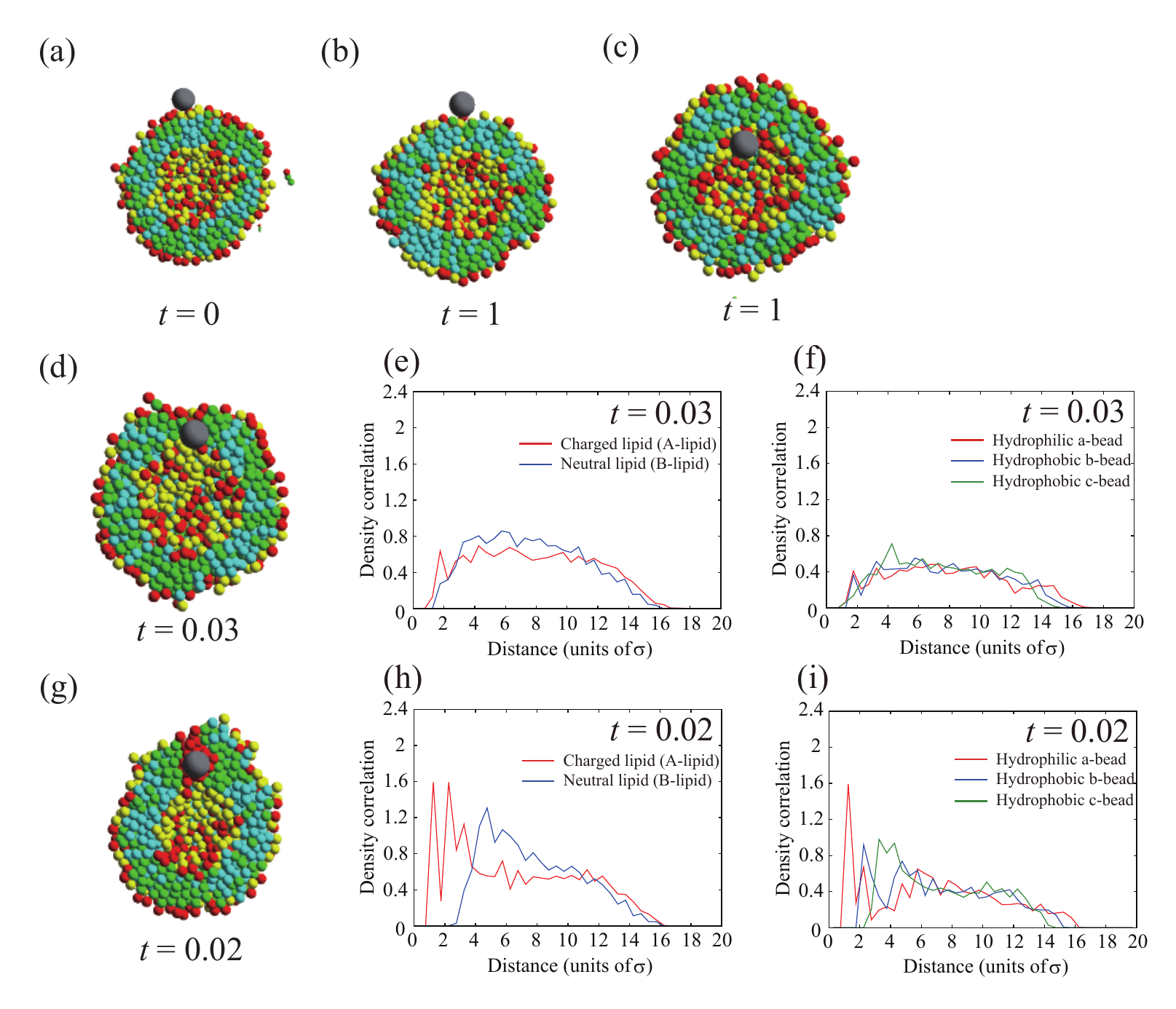}
\end{center}
\caption{
(a) Snapshot of initial state ($t=0 \times 7500 \tau$) for a vesicle composed of 500 
charged lipids (A-lipids) and 500 neutral lipids (B-lipids) with an NP ($d/\sigma = 2$). 
Charged lipid (A-lipid) is represented by red hydrophilic beads and green hydrophobic beads, while neutral lipid (B-lipid) is represented by yellow hydrophilic beads and cyan hydrophobic beads.
[(b) and (c)] Typical snapshots at $t=1 \times 7500 \tau$.
(b) Adsorption onto outer leaflet and (c) adsorption onto inner leaflet. 
(d) Typical snapshot when a neutral NP ($z_{\rm{n}} = 0$) passes through the hydrophobic 
core at $t=0.03 \times 7500 \tau$ for $d/\sigma = 5$, $v_{\rm nh}/k_{\rm B}T = 1$, and 
$v_{\rm nt}/k_{\rm B}T = 2$. 
[(e) and (f)] Density correlation functions of lipids and beads around the NP at $t=0.03 \times 7500 \tau$
corresponding to (d), respectively. 
Red and blue lines in (e) indicate the density correlation functions of A- and B-lipids, respectively. 
Red, blue, and green lines in (f) indicate the density correlation functions of a-, b-, and c-beads, 
respectively. 
(g) Typical snapshot when a positively charged NP ($z_{\rm {n}} = +150$) passes through a transient 
channel at $t=0.02 \times 7500 \tau$ for $d/\sigma = 2$, $v_{\rm nh}/k_{\rm B}T = 1$, and 
$v_{\rm nt}/k_{\rm B}T = 1$. 
[(h) and (i)] Density correlation functions of lipids and beads around the bead at $t=0.02 \times 7500 \tau$
corresponding to (g), respectively.}
\end{figure}

\clearpage
\subsection*{Data reproducibility}
Although we performed three calculations at least for each condition to ensure data reproducibility, 
we show a representative result in the calculations we performed in main text.
Here, we show the two other calculation results, and these two calculations will be referred to as 2nd calculation and 3rd calculation for convenience.

\begin{figure}[h!]
\begin{center}
\includegraphics{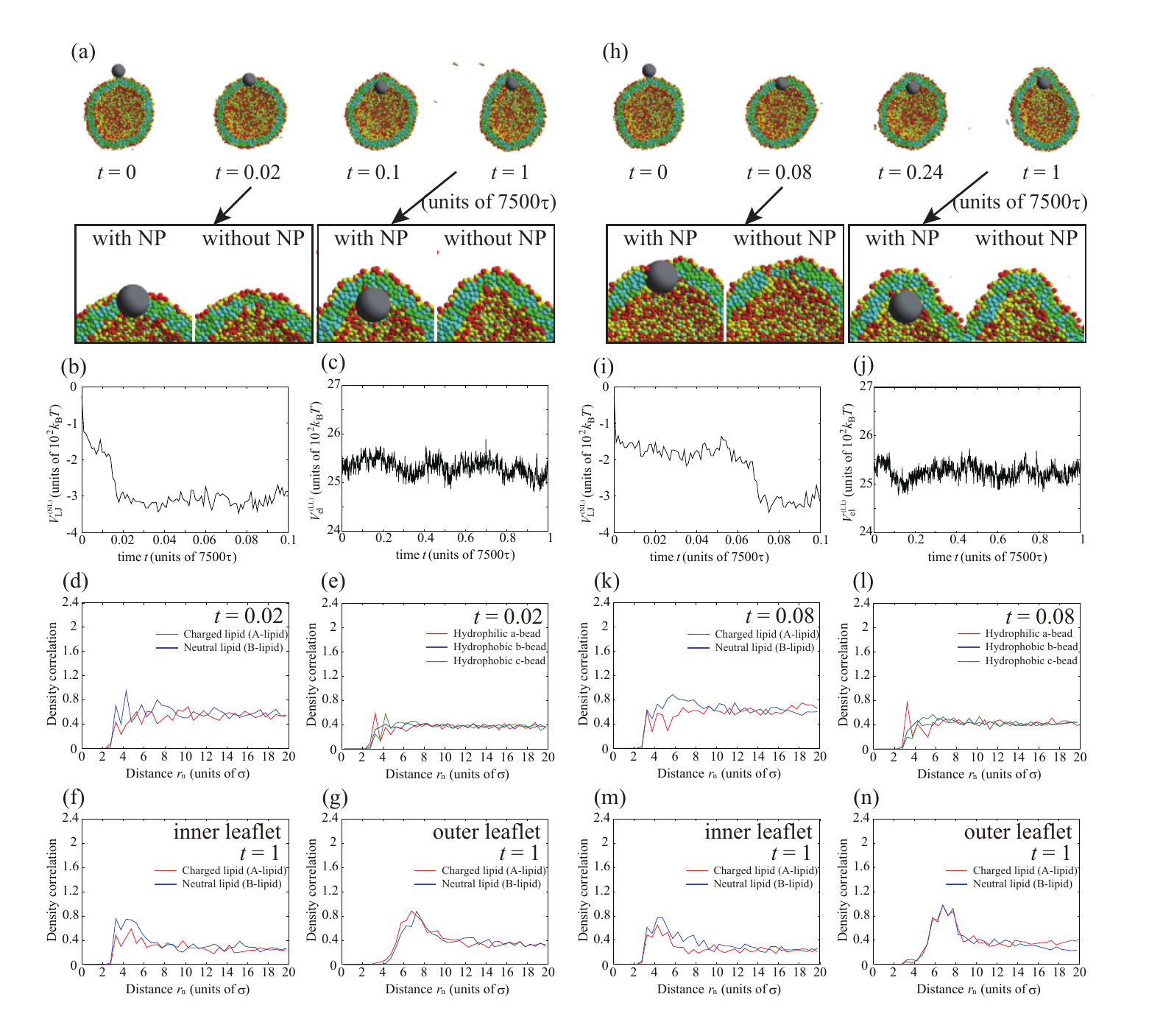}
\end{center}
\caption{
Permeation sequence of a neutral NP ($z_{\rm n}=0$) for $d/\sigma=5, v_{\rm nh}/k_{\rm B}T=2$, and $v_{\rm nt}/k_{\rm B}T=3$ of (a)-(g) 2nd and (h)-(n) 3rd calculations.
These calculations are performed at the same condition as Fig.1 to ensure data reproducibility. 
[(a) and (h)] Typical snapshots of the permeation sequence.
[(b) and (i)] Enlarged graph for time evolution of the Lennard-Jones potential between the NP and lipids until $t=0.1 \times 7500 \tau$.
[(c) and (j)] Time evolution of the electrostatic interaction between lipids.
[(d) and (k)] Density correlation functions of lipids around the NP.
[(e) and (l)] Density correlation functions of beads around the NP.
[(f), (g), (m), and (n)] Density correlation functions of A- and B-lipids in the inner leaflet and outer leaflet around the NP.
}
\end{figure}

\clearpage
\begin{figure}[h!]
\begin{center}
\includegraphics{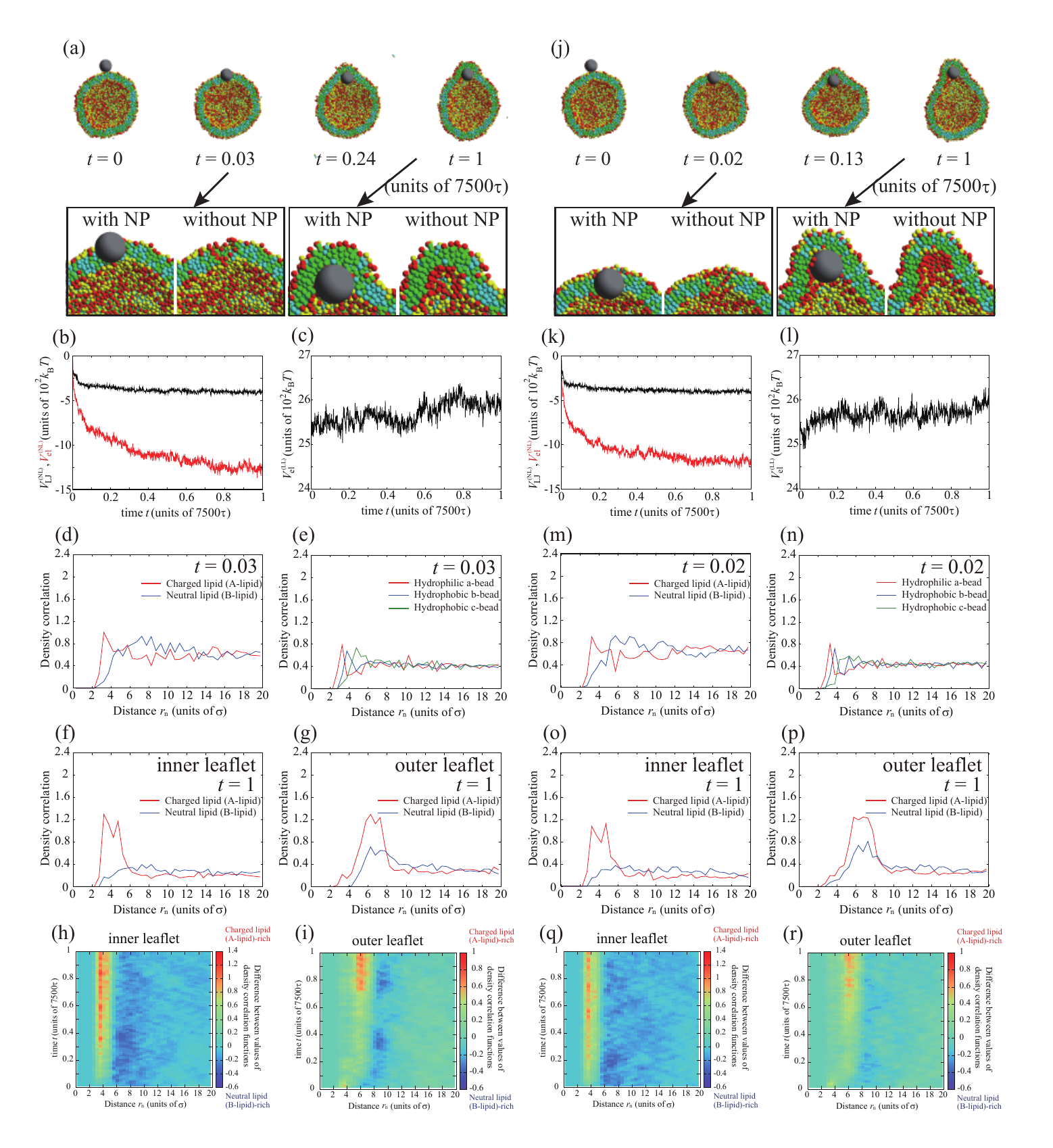}
\end{center}
\caption{
Permeation sequence of a charged NP ($z_{\rm n}=+50$) for $d/\sigma=5, v_{\rm nh}/k_{\rm B}T=2$, and $v_{\rm nt}/k_{\rm B}T=3$ of (a)-(i) 2nd and (j)-(r) 3rd calculations.
These calculations are performed at the same condition as Fig.3 to ensure data reproducibility. 
[(a) and (j)] Typical snapshots of the permeation sequence.
[(b) and (k)] Time evolution of the Lennard-Jones potential between the NP and lipids and the electrostatic interaction between the NP and lipids.
[(c) and (l)] Time evolution of the electrostatic interaction between lipids.
[(d) and (m)] Density correlation functions of lipids around the NP.
[(e) and (n)] Density correlation functions of beads around the NP.
[(f), (g), (o), and (p)] Density correlation functions of A- and B-lipids in the inner leaflet and outer leaflet around the NP at $t=1 \times 7500\tau$.
[(h), (i), (q), and (r)] Lipid distribution changes in the inner leaflet and outer leaflet as functions of distance and time.
}
\end{figure}

\clearpage
\begin{figure}[h!]
\begin{center}
\includegraphics{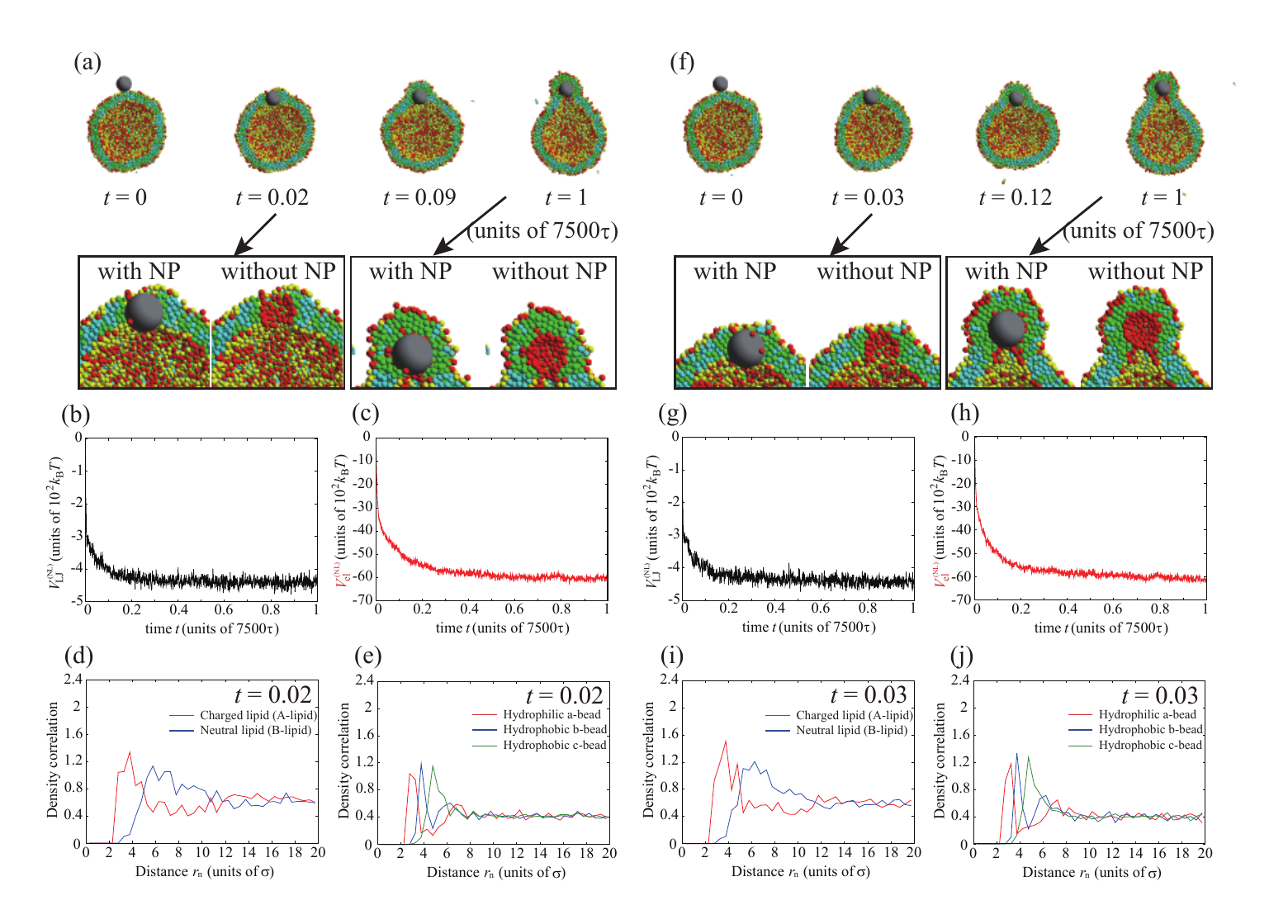}
\end{center}
\caption{
Permeation sequence of a charged NP ($z_{\rm n}=+150$) through a transient channel for $d/\sigma=5, v_{\rm nh}/k_{\rm B}T=2$, and $v_{\rm nt}/k_{\rm B}T=3$ of (a)-(e) 2nd and (f)-(j) 3rd calculations.
These calculations are performed at the same condition as Fig.4 to ensure data reproducibility. 
[(a) and (f)] Typical snapshots of the permeation sequence.
[(b) and (g)] Time evolution of the Lennard-Jones potential between the NP and lipids.
[(c) and (h)] Time evolution of the electrostatic interaction between the NP and lipids.
[(d) and (i)] Density correlation functions of lipids around the NP.
[(e) and (j)] Density correlation functions of beads around the NP.
}
\end{figure}

\clearpage
\begin{figure}[h!]
\begin{center}
\includegraphics{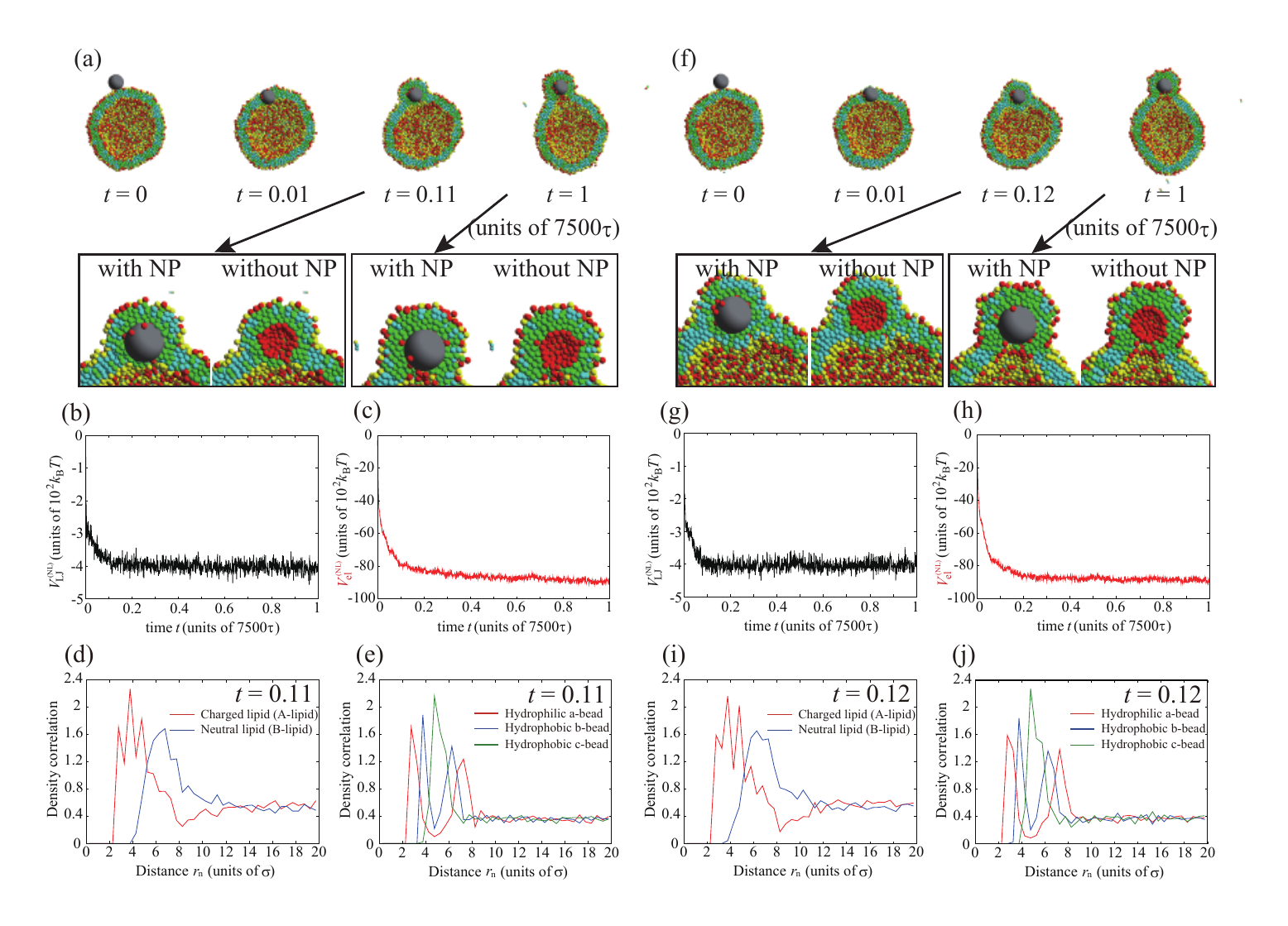}
\end{center}
\caption{
Permeation sequence of a charged NP ($z_{\rm n}=+200$) via transportlike endocytosis when $d/\sigma=5, v_{\rm nh}/k_{\rm B}T=2$, and $v_{\rm nt}/k_{\rm B}T=3$ of (a)-(e) 2nd and (f)-(j) 3rd calculations.
These calculations are performed at the same condition as Fig.5 to ensure data reproducibility. 
[(a) and (f)] Typical snapshots of the permeation sequence.
[(b) and (g)] Time evolution of the Lennard-Jones potential between the NP and lipids.
[(c) and (h)] Time evolution of the electrostatic interaction between the NP and lipids.
[(d) and (i)] Density correlation functions of lipids around the NP.
[(e) and (j)] Density correlation functions of beads around the NP.
}
\end{figure}

\clearpage
\begin{figure}[h!]
\begin{center}
\includegraphics{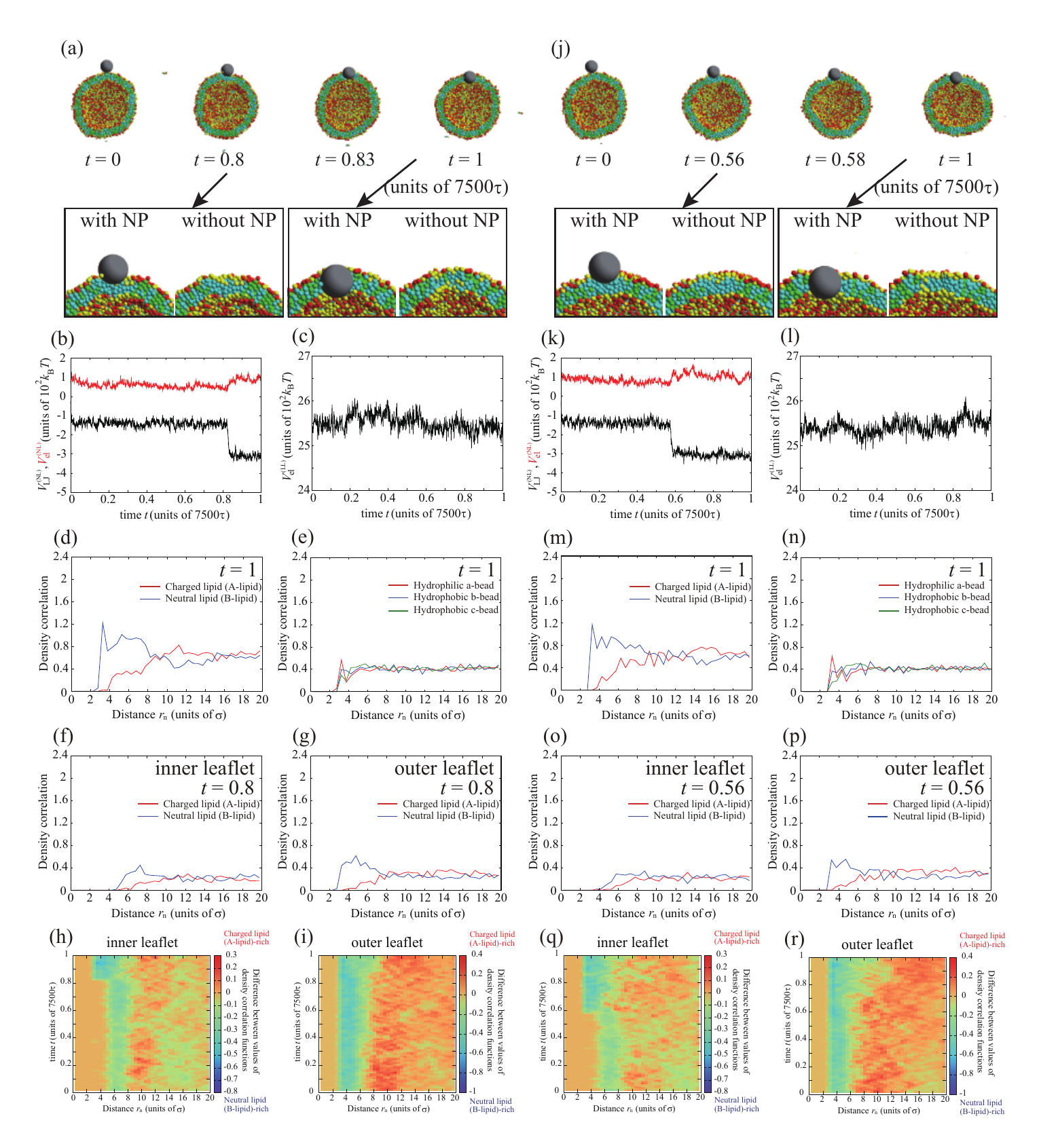}
\end{center}
\caption{
Behavior of a negatively charged NP ($z_{\rm n}=-50$) for $d/\sigma=5, v_{\rm nh}/k_{\rm B}T=2$, and $v_{\rm nt}/k_{\rm B}T=3$ of (a)-(i) 2nd and (j)-(r) 3rd calculations.
These calculations are performed at the same condition as Fig.6 to ensure data reproducibility. 
[(a) and (j)] Typical snapshots of negatively charged NP behavior.
[(b) and (k)] Time evolution of the Lennard-Jones potential between the NP and lipids and the electrostatic interaction between the NP and lipid.
[(c) and (l)] Time evolution of the electrostatic interaction between lipids.
[(d) and (m)] Density correlation functions of lipids around the NP.
[(e) and (n)] Density correlation functions of beads around the NP.
[(f), (g), (o), and (p)] Density correlation functions of A- and B-lipids in the inner leaflet and outer leaflet around the NP.
[(h), (i), (q), and (r)] Lipid distribution changes in the inner leaflet and outer leaflet as functions of distance and time.
}
\end{figure}

\clearpage
\subsection*{Adsorption angle}
\begin{figure}[h!]
\begin{center}
\includegraphics{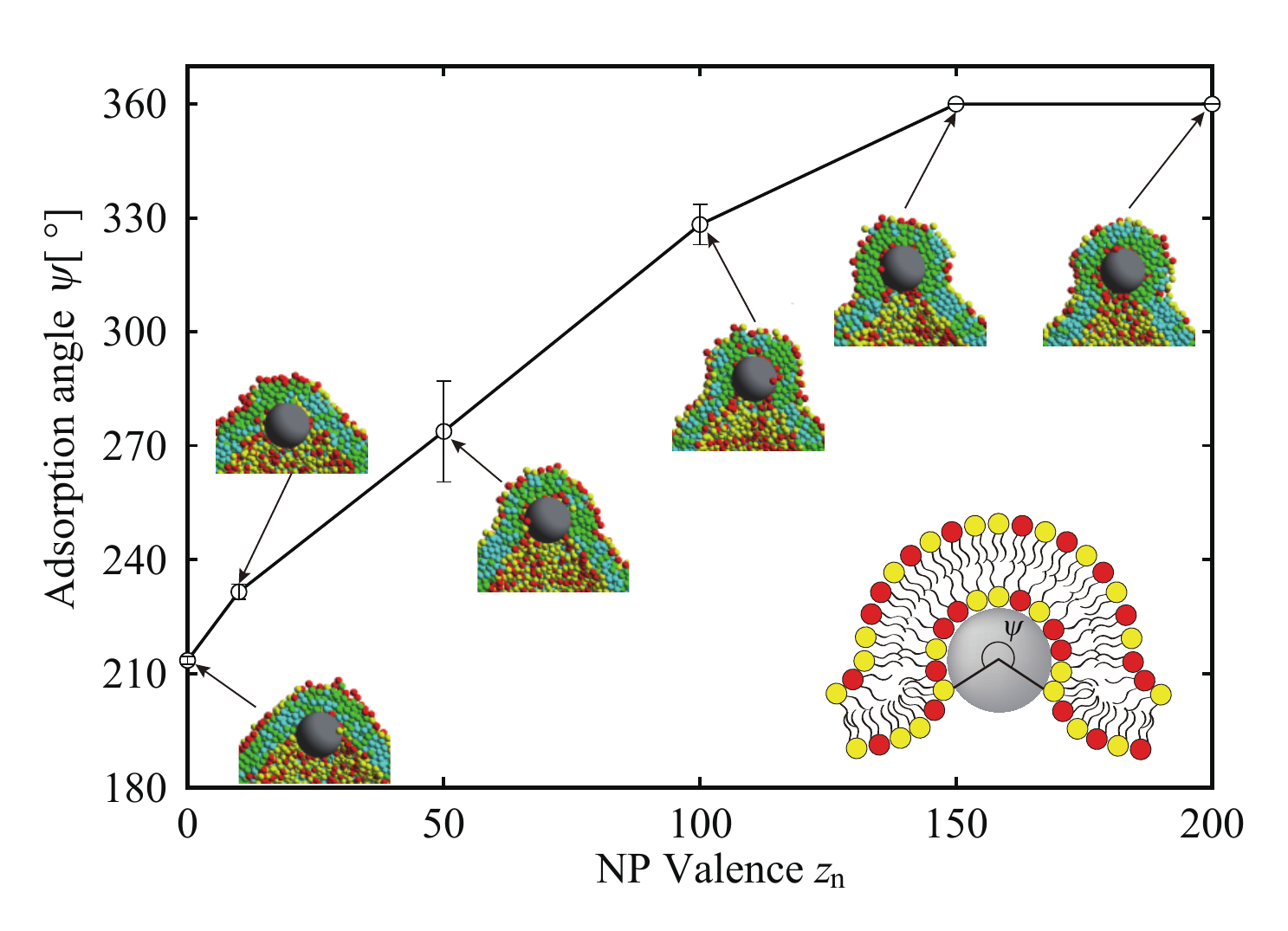}
\end{center}
\caption{
Adsorption angle $\psi$ which determines the adsorption area at $t=1\times 7500\tau$ as a function of the NP valence $z_{\rm n}$ for $d/\sigma=5, v_{\rm nh}/k_{\rm B}T=2$, and $v_{\rm nt}/k_{\rm B}T=3$.
Adsorption angle $\psi$ is represented by the schematic illustration.
Open circles are the average values of three calculations, and bars represent standard deviation.
}
\end{figure}

\end{document}